\documentclass{article}

\PassOptionsToPackage{numbers, compress}{natbib}

\usepackage[preprint]{neurips_2023}

\usepackage[utf8]{inputenc} 
\usepackage[T1]{fontenc}    
\usepackage{hyperref}       
\usepackage{url}            
\usepackage{booktabs}       
\usepackage{amsfonts}       
\usepackage{nicefrac}       
\usepackage{microtype}      
\usepackage{xcolor}         

\usepackage{wrapfig}
\usepackage{amssymb}
\usepackage{times}
\usepackage{latexsym}
\usepackage{booktabs}
\usepackage{multirow}
\usepackage{colortbl} 
\usepackage{graphicx}
\usepackage{rotating}
\usepackage{nicematrix}
\usepackage{enumitem}
\usepackage{minitoc}
\usepackage{tocloft}
\usepackage{sidecap} 
\usepackage{fontawesome}
\usepackage{pifont}
\usepackage{tikz}
\usepackage{rotating}
\usepackage{tcolorbox}
\usepackage{authblk}
\newcolumntype{C}[1]{>{\centering\arraybackslash}p{#1}}

\newenvironment{rotatepage}%
    {\clearpage\pagebreak[4]\global\pdfpageattr\expandafter{\the\pdfpageattr/Rotate 90}}%
    {\clearpage\pagebreak[4]\global\pdfpageattr\expandafter{\the\pdfpageattr/Rotate 0}}%

\newcommand*\emptycirc[1][1ex]{\tikz\draw (0,0) circle (#1);} 
\newcommand*\halfcirc[1][1ex]{%
  \begin{tikzpicture}
  \draw[fill] (0,0)-- (90:#1) arc (90:270:#1) -- cycle ;
  \draw (0,0) circle (#1);
  \end{tikzpicture}}
\newcommand*\fullcirc[1][1ex]{\tikz\fill (0,0) circle (#1);} 

\definecolor{deepred}{rgb}{0.631,0.102,0.102}
\definecolor{mildyellow}{HTML}{FFF2CC}
\definecolor{lightgray}{gray}{0.9}
\definecolor{colorEthical}{RGB}{255,242,204}
\definecolor{colorUnethical}{RGB}{255,204,204}

\hypersetup{
  colorlinks=true,
  linkcolor=black,      
  citecolor=blue,     
  urlcolor=deepred         
}

\definecolor{oaigreen}{HTML}{00a17a}
\usepackage{booktabs} 
\usepackage{subcaption}

\usepackage{multicol} 

\newenvironment{packedenumerate}{
\begin{enumerate}[label=(\arabic*), leftmargin=*]
\setlength{\itemsep}{0pt}
\setlength{\parskip}{0pt}
}{
\end{enumerate}
}

\newenvironment{takeaway}[1][]
  {
    \begin{tcolorbox}[%
        boxrule=0.5pt,
        arc=4pt,
        left=2pt,
        right=2pt,
        bottom=2pt,
        top=2pt,
        rounded corners
        ]
    \textbf{#1.}
    \small \itshape
    \begin{itemize}[leftmargin=1.3em,topsep=1pt,noitemsep]
  }
  {\end{itemize}\end{tcolorbox}}

\usepackage{fdsymbol}

\usepackage{pifont}
\definecolor{myblue}{HTML}{abc6ff}
\definecolor{myyellow}{HTML}{fbbc05}
\definecolor{myred}{HTML}{ea4335}

\title{AI Risk Categorization Decoded (AIR 2024): 
\\From Government Regulations to Corporate Policies}

\author{
\textbf{Yi Zeng}\textsuperscript{* 1,2} \quad
\textbf{Kevin Klyman}\textsuperscript{* 3,4} \quad
\textbf{Andy Zhou}\textsuperscript{5,6} \quad
\textbf{Yu Yang}\textsuperscript{1,7} \quad 
\textbf{Minzhou Pan}\textsuperscript{1,8} \quad 
\\
\textbf{Ruoxi Jia}\textsuperscript{2} \quad
\textbf{Dawn Song}\textsuperscript{1,9} \quad
\textbf{Percy Liang}\textsuperscript{3} \quad
\textbf{Bo Li}\textsuperscript{1,10} \quad
}

\affil{
\textsuperscript{1} \includegraphics[height=7.5pt]{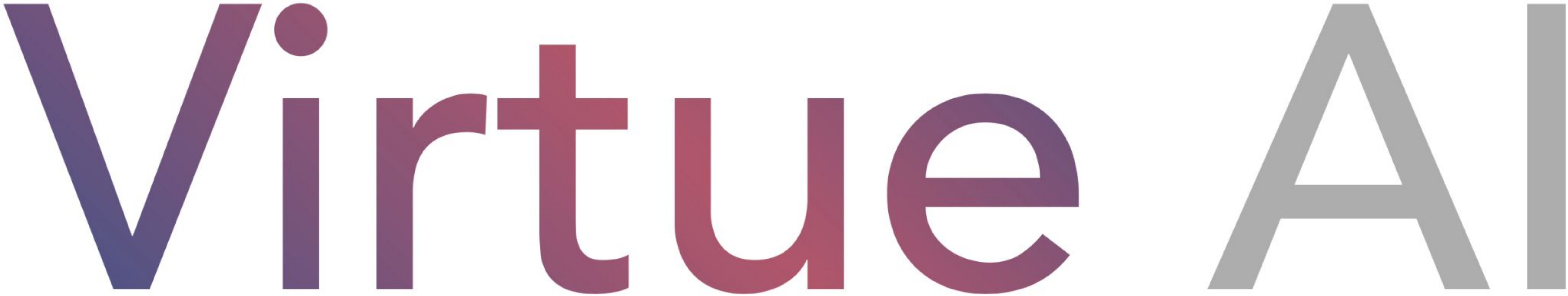}
\quad
\textsuperscript{2}Virginia Tech \quad
\textsuperscript{3}Stanford University \quad
\textsuperscript{4}Harvard University \quad
\textsuperscript{5}Lapis Labs \quad
\\
\textsuperscript{6}University of Illinois Urbana-Champaign \quad
\textsuperscript{7}University of California, Los Angeles \quad
\\
\textsuperscript{8}Northeastern University \quad
\textsuperscript{9}University of California, Berkeley \quad
\textsuperscript{10}University of Chicago \quad
}

\begin{document}

\maketitle

\begin{abstract}
We present a comprehensive AI risk taxonomy derived from 
eight government policies from the European Union, United States, and China and 16 company policies worldwide, making a significant step towards establishing a unified language for generative AI safety evaluation. 
We identify 314 unique risk categories, organized into a four-tiered taxonomy.
At the highest level, this taxonomy encompasses \textit{System \& Operational Risks}, \textit{Content Safety Risks}, \textit{Societal Risks}, and \textit{Legal \& Rights Risks}. 
The taxonomy establishes connections between various descriptions and approaches to risk, highlighting the overlaps and discrepancies between public and private sector conceptions of risk. 
By providing this unified framework, we aim to advance AI safety through information sharing across sectors and the promotion of best practices in risk mitigation for generative AI models and systems.
\end{abstract}

\begin{figure}[h!]
    \centering
    \vspace{-0.5em}
    \includegraphics[width=\linewidth]{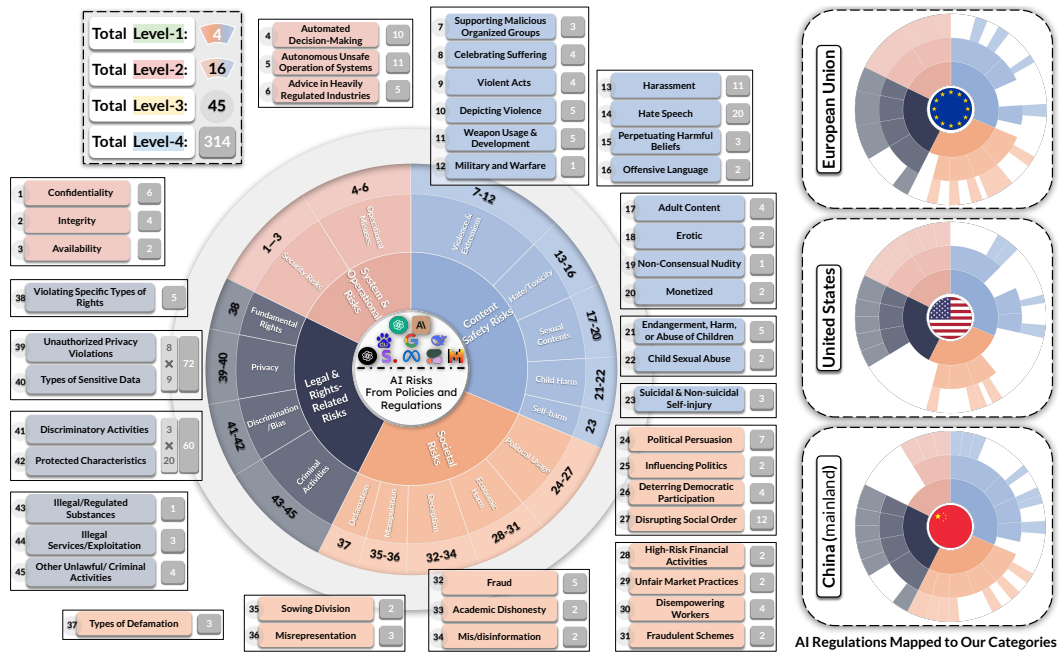}
    \caption{Overview of the AI risk taxonomy derived from 24 policy and regulatory documents, encompassing 314 unique risk categories. Charts on the right-hand side map to major AI regulations.}
    \label{fig:overview}
\end{figure}

\newcommand{\pl}[1]{\textcolor{red}{[PL: #1]}}

\tableofcontents
\newpage


\hypersetup{
  colorlinks=true,
  linkcolor=deepred,      
  citecolor=blue,     
  urlcolor=deepred         
}

\section{Introduction}
\label{sec:intro}

The rapid integration of foundation models~\citep{chatgpt,gpt4v,openai2023gpt4,touvron2023llama,touvron2023llama-2,claude,geminiteam2023gemini} into various sectors of the economy has highlighted the immense potential of general-purpose AI. 
However, the broad capabilities of foundation models introduce a complex spectrum of new risks while reinforcing existing threats. 
Governments and companies have responded quickly, implementing regulations and policies to address the risks from AI~\citep{eu-ai-act,EOWhiteHouse,china-recomandations,china-synthesis,china-genai}; in parallel, academic researchers have also explored and proposed numerous AI safety benchmarks and taxonomies of the risks from AI ~\citep{gehman2020realtoxicityprompts,wang2023decodingtrust,qi2024finetuning,li2024salad,chao2024jailbreakbench,zou2023universal,mazeika2024harmbench}. 

These regulations and policies are often siloed. 
Despite ongoing efforts, there is no unified categorization of AI risks that comprehensively covers all domains of risk while taking into account the perspective of industry and government. 
Academic benchmarks primarily rely on existing literature, often failing to fully incorporate the latest government frameworks and company policies. 
Companies at the forefront of the development and deployment of foundation models have policies that reflect their understanding of potential risks, but these policies are tailored to the laws of the jurisdictions in which they operate. 
Government regulations and policies list high-level risks that prioritize societal concerns, but often lack the granularity to address lower-level risks, such as the potential for large language models to be used
to promote self-harm---a concern highlighted in company policies and academic research. 
The development of independent categorizations of AI risk within each sector can lead to an incomplete understanding of the full risk landscape, ultimately hindering the safe deployment of foundation models.

This paper proposes the AI Risk Taxonomy (\textbf{AIR 2024}), a unified taxonomy of risks that addresses gaps across different companies and jurisdictions. 
Unlike existing risk taxonomies, AIR 2024 is grounded in government regulation and company policies. 
This approach ensures relevance and applicability across jurisdictions while providing a cohesive framework for integrating diverse efforts across sectors and regions. 
Our contributions are as follows:
\begin{packedenumerate}
\item \textbf{Unified AI Risk Taxonomy}: AIR 2024, informed by policies from AI companies as well as the EU, US, and China (mainland), identifies 314 risk types
and structures them into a four-level hierarchy. This standardized framework enables AI safety evaluations and provides a basis for consistent assessment of AI-related risks in different regions.
\item \textbf{Private Sector Risk Categorization Analysis} (\S\ref{sec:policies}): Using AIR 2024, we analyze how companies categorize AI risks, providing insights into how organizations that develop and deploy generative AI models perceive and prioritize these concerns. This analysis identifies trends, biases, and gaps in current corporate risk management strategies.
\item \textbf{AI Regulatory Risk Categorization Analysis} (\S\ref{sec:regulations}): Based on AIR 2024, we conduct a comparative analysis of AI regulations from the EU, US, and China, highlighting similarities and differences in regional AI governance approaches and contributing to a deeper understanding of how legislative landscapes influence the development and deployment of generative AI systems.
\item \textbf{Discussion and Case Study} (\S\ref{sec:iterplays}): We assess the agreement between corporate and government policies by considering the case of Chinese companies, offering practical insights into how company practices align with or diverge from existing government regulations. We also provide takeaways from this work and highlight areas for future research.
\end{packedenumerate}
AIR 2024 harmonizes terms
across industry and government contexts, facilitating a uniform understanding of AI risks. 
This uniformity is crucial for companies and academics operating internationally, where disparate regulations can lead to confusion and compliance challenges. 
By identifying shared risk categories, our methodology yields a common language for clearer communication and more effective collaboration among policymakers, industry leaders, academic researchers, and regulatory bodies.
Our comparative analysis between government and company policies highlights areas where regulatory frameworks might be underdeveloped, suggesting focus areas for policymakers. 
It also identifies areas where company policies are more stringent or advanced than regulations, pointing to potential best practices that could inform future regulatory efforts.
The insights derived from our line-by-line analysis of policies provide a data-driven baseline for further policy development. 
Moreover, these insights contribute to the responsible development and deployment of generative AI systems, promoting safety, fairness, and transparency within the industry.

\begin{rotatepage} 
\begin{sidewaysfigure}
    \centering
    \includegraphics[width=1\linewidth]{figs/AIR_2024_new.pdf}
    \caption{\textbf{The AIR Taxonomy, 2024}: The complete set of 314 structured risk categories spanning four levels: \scalebox{0.9}{\colorbox[HTML]{DAEBD3}{\textbf{level-1}}} consists of four general high-level categories; \scalebox{0.9}{\colorbox[HTML]{F4CDCC}{\textbf{level-2}}} groups risks based on societal impact; \scalebox{0.9}{\colorbox[HTML]{FFF3CC}{\textbf{level-3}}} further expands these groups; \scalebox{0.9}{\colorbox[HTML]{CFE3F4}{\textbf{level-4}}} contains detailed risks explicitly referenced in policies and regulations.
    }
    \label{fig:air_2024}
\end{sidewaysfigure}
\end{rotatepage}

\global\pdfpageattr\expandafter{\the\pdfpageattr/Rotate 0}

\section{Methodology}
\label{sec:method}


Recognizing the that existing AI risk taxonomies \citep{weidinger2021ethical, klyman2024aups-for-fms, wang2023decodingtrust} are not fully reflective of corporate policies and government regulations, we propose a systematic, bottom-up approach to construct an AI risk taxonomy grounded in public and private sector policies.
Whereas other taxonomies of the risks and harms of generative AI models and systems draw primarily on existing literature \citep{weidinger2023sociotechnical, shelby2023sociotechnical, Hoffmann2023}, we taxonomize risk based on how companies and governments describe risks in their own policies. 
As in \citep{klyman2024aups-for-fms}, we used a qualitative content analysis to code the risk categories in policies from governments and companies \citep{Mayring2015}. 
This was done inductively \citep{https://doi.org/10.1111/j.1365-2648.2007.04569.x}, with categories drawn directly from such policies. 
The process of constructing the AIR 2024 involved the following steps:

\begin{packedenumerate}
    \item \textbf{Collection of Policies}: We begin by collecting a diverse set of policies, focusing on their relevance, comprehensiveness, and diversity. In total, this version of the taxonomy covers the risk categories specified by eight government policies from the European Union, the United States, and China, as well as 16 company policies from nine leading foundation model developers selected for their comprehensive specification of risk categories. 
    We focus on government policies that include some binding restrictions on generative AI models and companies' acceptable use policies.
    We provide the detailed collection of company policies in Figure \ref{tab:policy_landscape} and government policies in Section \ref{sec:regulations}, respectively.

    \item \textbf{Risk Extraction}: We analyze each policy and regulation using a consistent process to extract and organize risk categories that are explicitly referenced in each policy document. 
    This involves parsing every line of each document, manually clustering related sections, identifying specific risks, and rephrasing them to capture overlap and maintain consistency while highlighting unique categories \citep{https://doi.org/10.1111/j.1365-2648.2007.04569.x}. 
    Throughout this process, we perform a comparative analysis of risk categories across different policies and regulations to identify similarities and differences in how various entities and jurisdictions address similar risks. 
    For example, when analyzing risks related to ``unqualified usage,'' we compare OpenAI's recently updated usage policies \cite{OpenAI_new} (which prohibit ``Providing tailored legal, medical/health, or financial advice without review by a qualified professional$\ldots$'') and Google's prohibited use policy for its Gemma model series \cite{Google_gemma} (which prohibits ``Engagement in unlicensed practices of any vocation or profession including, but not limited to, legal, medical, accounting, or financial professional'' and ``Misleading claims of expertise or capability made particularly in sensitive areas (e.g. health, finance, government services, or legal)''). We identify shared categories of risks related to language models providing advice in legal, medical, and financial services, despite slight differences in the phrasing of the policies. 
   As another example, the Gemma prohibited use policy includes risks related to the use of the model in accounting and government services, which are two unique risk categories that do not appear in the policies of other foundation model developers.

    \item \textbf{Taxonomy Construction}: The risks we extract are organized into a hierarchical taxonomy using a bottom-up approach. 
    Granular risks that are described in detail (such as the example above) are mapped to level-4 categories, which are then grouped into broader level-3 and level-2 categories based on their similarity and the context in which they are referenced in policies. 
    For instance, the level-3 risk of ``advice in heavily regulated industries'' is grouped with ``automated decision making'' and ``autonomous unsafe operation of systems'' to form the level-2 category ``Operational Misuses,'' capturing the overarching theme of risks due to certain autonomous risks. The level-2 categories are further aggregated into four level-1 categories: ``System \& Operational Risks,'' ``Content Safety Risks,'' ``Societal Risks,'' and ``Legal \& Rights-Related Risks,'' as illustrated in Figure \ref{fig:overview}.  
\end{packedenumerate}

This result of this process is a work in progress. 
Many of the government policies we consider have yet to take full effect.
For example, China is in the process of finalizing the implementing 
regulations for its Interim Measures for the Management of Generative Artificial Intelligence Services \citep{China_Safety_AI_2024}. 
The Codes of Practice that will determine how much of the EU AI Act is enforced have yet to be drafted \citep{hacker2023ai}. 
And the extent to which the US Executive Order on the Safe, Secure, and Trustworthy Development and Use of Artificial Intelligence has been implemented remains opaque \citep{Meinhardt2024transparency}.
Companies regularly change their policies, as evidenced by the shift in OpenAI's Usage Policies that we document. 
We intend to update this taxonomy as government and company policies evolve. 
Nevertheless, these major AI regulations have been adopted and have significant bearing on how companies and government agencies conceive of and address risk sfrom AI.

During the development of this taxonomy, we encountered significant challenges due to the diversity of provisions within different policies across organizations. 
Companies and governments use to different terminology to describe similar topics, presenting a potential for inconsistency. 
To address this issue and ensure consistency, we adhered to the three-step process above while constructing the AIR 2024. 
Additionally, to avoid inaccuracies and errors that might arise from language model hallucination, we deliberately refrained from employing language models or summarization tools in our process of categorizing and analyzing risks.

The complete list of the 314 risk categories identified through our method is presented in Figure \ref{fig:air_2024}, which provides a comprehensive mapping of the AI risk landscape by integrating granular terms referenced in current regulatory frameworks and industry policies.
Risks are color-coded according to their position in our hierarchical taxonomy: \scalebox{0.9}{\colorbox[HTML]{DAEBD3}{\textbf{level-1}}} (total 4), \scalebox{0.9}{\colorbox[HTML]{F4CDCC}{\textbf{level-2}}} (total 16), \scalebox{0.9}{\colorbox[HTML]{FFF3CC}{\textbf{level-3}}} (total 45), and \scalebox{0.9}{\colorbox[HTML]{CFE3F4}{\textbf{level-4}}} (total 314). 
For clarity, when referring to a specific risk category in our taxonomy in this paper, we use color coding to indicate its level in the taxonomy.

\section{Private Sector Categorizations of Risk}
\label{sec:policies}

This section presents a risk taxonomy drawn from 16 policies of 9 foundation model developers (Figure \ref{fig:air_2024}). 
We focus on two types of company policies that seek to govern generative AI in order to address specific risks: \textbf{platform-wide acceptable use policies} and \textbf{model-specific acceptable use policies} \citep{klyman2024aups-for-fms}. 
An overview of the company policies we consider in this study organized into 13 sets is listed in Table \ref{tab:policy_landscape}.

\begin{table*}[h]
\centering
\includegraphics[width=\linewidth]{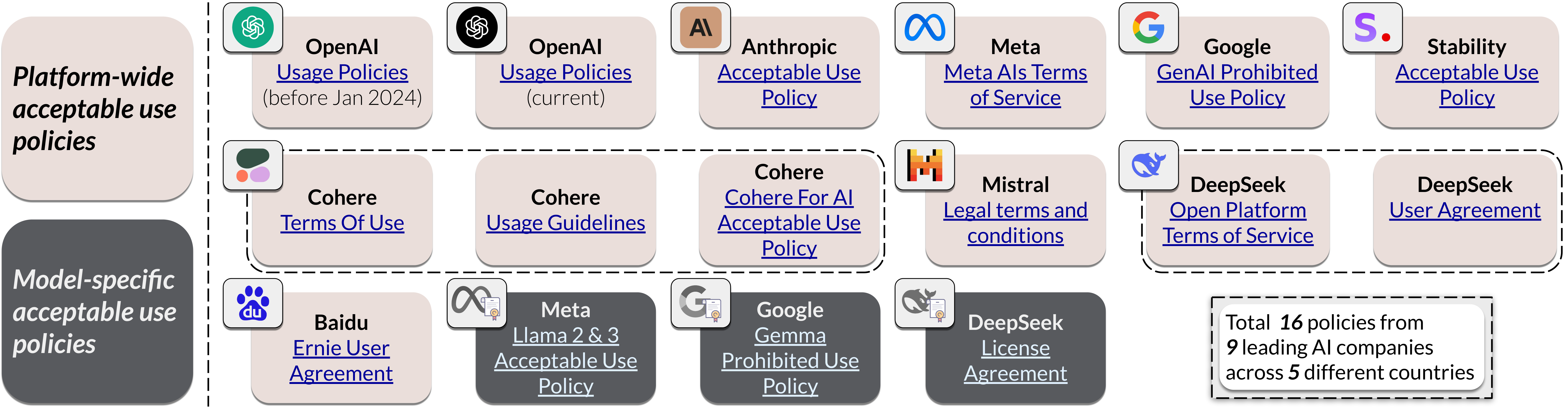}
\caption{Overview of the company policies (16 documents organized into 13 sets) we consider in this study.}
\label{tab:policy_landscape}
\end{table*}

\textbf{Platform-wide acceptable use policies} include documents labeled as terms of service and usage guidelines~\citep{klyman2024aups-for-fms}, which define categories of risky use that are restricted or prohibited across a company's products, services, and platforms. 
We analyze a diverse range of policies from leading AI firms across different countries, providing a comprehensive set of policies detailing the uses of their generative AI models and systems that they prohibit. 
The platform-wide policies in this study include the 2023 and 2024 versions of OpenAI's usage policies~\citep{OpenAI_old, OpenAI_new}, Anthropic's acceptable use policy~\citep{Anthropic_aup}, Meta AI's terms of service~\citep{Meta_ai}, Google's prohibited use policy~\citep{Google_genai}, Cohere For AI's acceptable use policy~\citep{Cohere_aup}, terms of use~\citep{Cohere_tos}, and usage guidelines~\citep{Cohere_ug}, Mistral's legal terms and conditions (encompassing terms of use, terms of service for La Plateforme, and terms of service for Le Chat)~\citep{Mistral_legal}, Stability's acceptable use policy~\citep{Stability_aup}, DeepSeek's open platform terms of service~\citep{DeepSeek_platform} and terms of use~\citep{DeepSeek_user}, and Baidu's user agreement for Ernie~\citep{Baidu_user}.

\textbf{Model-specific acceptable use policies} are tied to specific open-source foundation models (i.e., models with publicly available weights) and serve as a primary means of governing their use \citep{kapoor2024societal, Contractor_2022}. 
We analyze license terms from prominent open-source models such as the acceptable use policy for Meta's Llama 2 and Llama 3 models \citep{Meta_llama2}, Google's prohibited use policy for Gemma \citep{Google_gemma}, and DeepSeek's license agreement for DeepSeek LLM \citep{DeepSeek_model}. 
It is necessary to distinguish between platform-wide policies and policies that are tailored to specific open models because many open foundation models are primarily deployed locally, meaning that model developers have no platform through which they can enforce their policies against most users \citep{downing2023licensing}.

\newpage
\textbf{We did not include the following policies in our study:}

\textit{Company policies that are too abstract and simplified}: Although other leading firms, such as Microsoft \citep{microsfot_tos}, 01.AI \citep{yi01ai}, Amazon \citep{aws_responsible}, and Alibaba \citep{qwen}, have contributed significantly to the AI ecosystem and AI safety landscape, their policies restricting particular uses of AI models are too general to aid in our analysis. 
For example, 01.AI's license for its Yi model series contains relatively few categories of prohibited use \citep{klyman2024aups-for-fms}. 
As these policies would not introduce new risk categories to supplement our taxonomy, we focus on more detailed policies, which offer more comprehensive risk analyses for comparison and analysis.

\textit{Other documents that only outline safety standards without specifying AI risk categorizations}: There are a number of industry guidelines \citep{Chohere}, checklists \citep{Owasp}, maturity models \citep{Salesforce,IBM}, and standards \citep{Microsoft,OpenAI_spec} that relate to AI and safety. 
However, many of these documents focus on defining the characteristics of a safe AI system or outlining general problems with machine learning models (e.g., trustworthiness, hallucination, or bias) without delineating specific risk categories relevant to downstream use. 
Similarly, we exclude Responsible Scaling Policies (or preparedness policies) \citep{Anthropic,openai2023preparedness} that guide a company's decision about whether to release a foundation model based on tracking its capabilities in specific high-risk areas (e.g., biorisk, cyber risk). 
Our aim is to primarily assess categories of risk that companies take steps to legally prohibit, as these risks are most directly comparable to binding prohibitions in government policies. 

\subsection{Unpacking the Risk Categories}

In this section, we present a mapping of risk categories specified by company policies to our final risk taxonomy at level-3. 
Table \ref{tab:each_companies} provides the main comparison of different companies and the percentage of risks specified in their policies covering our taxonomy at level-3 risk categories. 
In comparison, DeepSeek, Anthropic, OpenAI, and Stability AI cover the largest number of risk categories, with all above 70\% coverage reflected on the level-3 categories in the AIR 2024. 
This coverage does not indicate the direct efforts of each company in their safety mitigation. Each company's policy is more tailored to the specific regime they are operating in. While DeepSeek has the most comprehensive coverage of risk categories, it is also the only company providing services to the European Union, the United States, and China. Other companies, on the other hand, provide services in at most two of these jurisdictions.
Moreover, additional coverage of risk categories is not necessarily a good thing.
For instance, Chinese regulators' efforts to force companies to avoid some of the risks referenced in their internal policies (e.g., ``subverting state power,'' ``damaging state interests,'' ``undermining national unity'') amount to censorship \citep{toner2023}.
While discussion of more granular risks is omitted here, the detailed risk categorization, from level-1 to level-4, is available in Figure \ref{fig:air_2024}.

\begin{table*}[h!]
    \centering
    \includegraphics[width=0.98\linewidth]{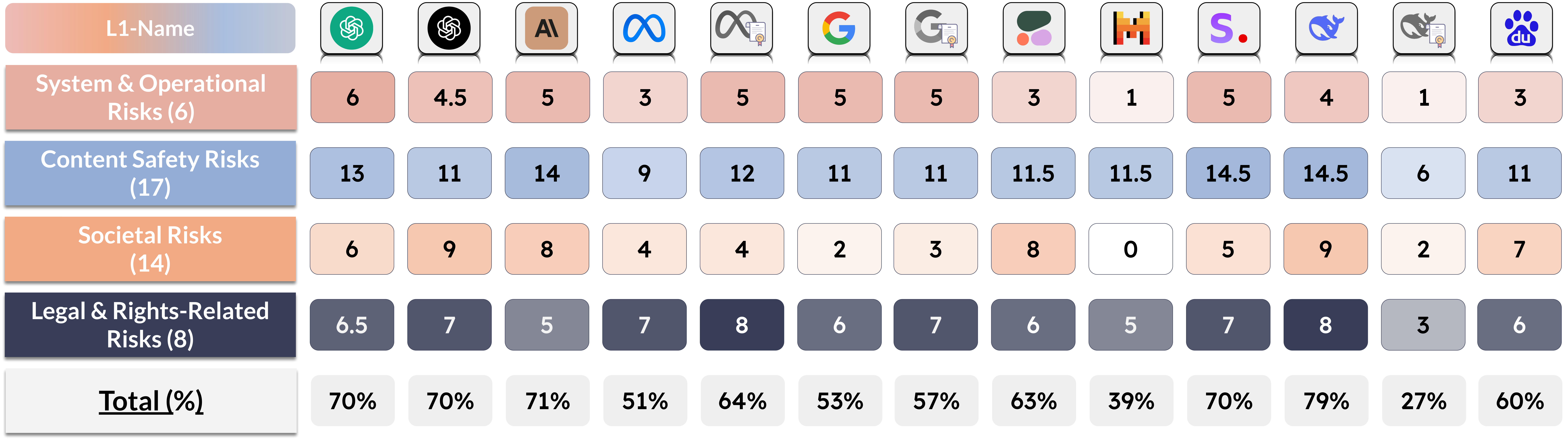}
    \caption{Risk categories covered by each company's policies at level-3 risks in our AIR Taxonomy. Categories that are referenced without further elaboration are counted as 0.5.}
    \label{tab:each_companies}
\end{table*}

This section details our analysis of each set of company policies with respect to the four level-1 categories in each subsection (i.e., \scalebox{0.9}{\colorbox[HTML]{DAEBD3}{\textbf{System \& Operational Risks}}}, \scalebox{0.9}{\colorbox[HTML]{DAEBD3}{\textbf{Content Safety Risks}}}, \scalebox{0.9}{\colorbox[HTML]{DAEBD3}{\textbf{Societal Risks}}}, and \scalebox{0.9}{\colorbox[HTML]{DAEBD3}{\textbf{Legal \& Rights-Related Risks}}}).

Each table in the following part of this section uses circles to indicate the depth and specificity of each policy's coverage: filled circles (\fullcirc[0.8ex]) represent explicit mentions of level-4 risk categories under that specific level-3 category, half-filled circles (\halfcirc[0.8ex]) denote brief mentions of general descriptions related to a specific level-3 category but without elaboration (e.g., level-2 descriptions), and empty circles (\emptycirc[0.8ex]) indicate an absence of any substantial language related to the specific risk category.




\subsubsection{
\scalebox{1}{\colorbox[HTML]{DAEBD3}{System \& Operational Risks}}}

\textbf{Overview.}
Table \ref{tab:level-1-systemic} presents a summary of the six level-3 risk categories within the level-1 category ``\scalebox{0.9}{\colorbox[HTML]{DAEBD3}{\textbf{System \& Operational Risks}}},'' comparing their coverage across 13 sets of different corporate policies denoted in Figure \ref{tab:policy_landscape}. 
The number of more granular level-4 risks that are explicitly referenced is listed alongside each level-3 risk category (there are a total of 38 such risks). 
These risks primarily concern the potential misuse of foundation models to compromise cybersecurity or as part of systems in highly regulated industries. 

\begin{table*}[h!]
    \centering
    \vspace{-0.5em}
    \includegraphics[width=\linewidth]{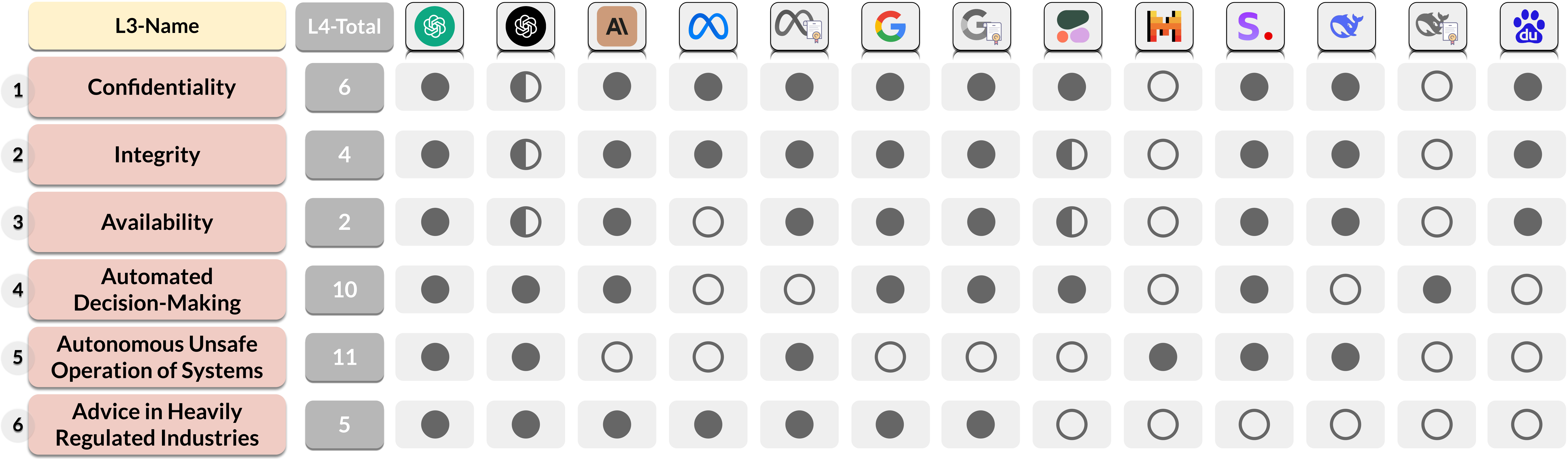}
    \caption{Corporate policy risk mapping: \textbf{A. \scalebox{0.9}{\colorbox[HTML]{DAEBD3}{\textbf{System \& Operational Risks}}}}. This level-1 risk category consists of two level-2 risk categories: \scalebox{0.9}{\colorbox[HTML]{F4CDCC}{\textbf{\textit{Security Risks}}}} and \scalebox{0.9}{\colorbox[HTML]{F4CDCC}{\textbf{\textit{Operational Misuse}}}}. These categories further break down into six level-3 categories shown in the figure and 38 level-4 risks.}
    \label{tab:level-1-systemic}
    \vspace{-0.5em}
\end{table*}

\textbf{Frequently and infrequently referenced categories.}
We observe that the categories of risks that fall under the level-2 category \scalebox{0.9}{\colorbox[HTML]{F4CDCC}{System Security}}---\scalebox{0.9}{\colorbox[HTML]{FFF3CC}{Confidentiality}}, \scalebox{0.9}{\colorbox[HTML]{FFF3CC}{Integrity}}, and \scalebox{0.9}{\colorbox[HTML]{FFF3CC}{Availability}}---are the risk categories that are most frequently referenced in model developers' policies, with all being referenced by more than 10 of the 13 sets of company policies; many company policies also include references to  level-4 risks in this area (e.g., \scalebox{0.9}{\colorbox[HTML]{CFE3F4}{Malware}}). 
Conversely, \scalebox{0.9}{\colorbox[HTML]{FFF3CC}{Autonomous Unsafe Operation of Systems}} receives less coverage, with only 6 of the 13 sets of company policies explicitly discussing risks relevant to this category. 
This disparity highlights a potential gap in addressing the unique challenges and risks associated with incorporating generative AI models into autonomous systems without a human in the loop.


\textbf{Comparative analysis.}
OpenAI's 2023 usage policy distinguishes itself by offering comprehensive and detailed coverage across all level-3 risk categories, accompanied by a substantial number of fine-grained level-4 risks. 
OpenAI's 2024 usage policies have a more simplified risk categorization that briefly mentions system security, indicating a transition from focused categorization to a more general approach.
In the case of Meta, its license for Llama 2 and Llama 3 is more detailed with respect to \scalebox{0.9}{\colorbox[HTML]{DAEBD3}{System \& Operational Risks}} than its platform-wide Terms of Service for its Meta AI service. 
Meanwhile, policies from Mistral and the model license from DeepSeek both focus on one specific risk among the 6 level-3 risks, suggesting a more narrow approach to risk categorization that may benefit from further refinement. 
Considering DeepSeek's model-specific policy and its platform-wide policies, its model license is more general than its platform-wide policy, indicating a different approach in comparison to Meta (with the model license being more specific) and Google's approach (with the platform and model-specific policies covering the same risks using the same language).

\begin{takeaway}[Takeaways]
\item Most company policies comprehensively detail risks related to security threats to other systems.
\item Risks associated with AI overreliance or excessive autonomy are less frequently specified in detail.
\item Companies with both platform-wide and model-specific policies vary in their approach to how they taxonomize risk in these different policy documents.
\end{takeaway}
%

\subsubsection{\scalebox{1}{\colorbox[HTML]{DAEBD3}{Content Safety Risks}}}


\textbf{Overview.}
Table \ref{tab:level-1-harmful} presents the 17 level-3 risk categories within the level-1 category of \scalebox{0.9}{\colorbox[HTML]{DAEBD3}{Content Safety Risks}} mapped to the 13 sets of companies' AI policies. 
This level-1 category consists of 79 unique level-4 risk categories. 
These risks primarily concern the direct harms associated with AI-generated, aiming to protect users from related to content safety, such as hate speech, harassment, and explicit material. 

\begin{table*}[h!]
\vspace{-.5em}
    \centering
    \includegraphics[width=\linewidth]{figs/Harmful-Content.pdf}
    \caption{Corporate policy risk mapping: \textbf{B. \scalebox{0.9}{\colorbox[HTML]{DAEBD3}{Content Safety Risks}}}. Risk categories identified under this level-1 risk consist of 5 level-2 risk categories: \scalebox{0.9}{\colorbox[HTML]{F4CDCC}{\textit{\textbf{Violence \& Extremism}}}}, \scalebox{0.9}{\colorbox[HTML]{F4CDCC}{\textit{\textbf{Hate/Toxicity}}}}, \scalebox{0.9}{\colorbox[HTML]{F4CDCC}{\textit{\textbf{Sexual Content}}}}, \scalebox{0.9}{\colorbox[HTML]{F4CDCC}{\textit{\textbf{Child Harm}}}}, and \scalebox{0.9}{\colorbox[HTML]{F4CDCC}{\textit{\textbf{Self-harm}}}}. The risk categories further break down into 17 level-3 categories shown and 79 unique level-4 categories.}
    \label{tab:level-1-harmful}
    \vspace{-.5em}
\end{table*}

\textbf{Frequently and infrequently referenced categories.} The level-3 categories \scalebox{0.9}{\colorbox[HTML]{FFF3CC}{Harassment}}, \scalebox{0.9}{\colorbox[HTML]{FFF3CC}{Celebrating}} \scalebox{0.9}{\colorbox[HTML]{FFF3CC}{Suffering}}, \scalebox{0.9}{\colorbox[HTML]{FFF3CC}{Monetized Sexual Content}}, and \scalebox{0.9}{\colorbox[HTML]{FFF3CC}{Child Sexual Abuse}} emerge as the most commonly referenced risk categories, with nearly all sets of policies (at least 12 of 13) providing detailed level-4 risks. 
This widespread coverage highlights the industry's recognition of the severe consequences of such types of AI misuse. On the other hand, \scalebox{0.9}{\colorbox[HTML]{FFF3CC}{Non-Consensual Nudity}} and \scalebox{0.9}{\colorbox[HTML]{FFF3CC}{Offensive Language}} receive comparatively less attention, with only 1 or 2 out of 13 sets of company policies explicitly specifying these categories. This disparity suggests that some content-related risks may be overlooked or considered less critical by certain companies.



\textbf{Comparative analysis.} Anthropic, Stability, and DeepSeek stand out for their comprehensive coverage of nearly all level-3 risk categories under this level-1 category, with each prohibiting a substantial number of granular level-4 risks. 
In contrast to its platform-wide policy, DeepSeek's model license exhibits a more focused approach, addressing only 5 out of 17 risk categories in detail while omitting others.
Comparing Stability's acceptable use policy to others, we notice a unique emphasis on the \scalebox{0.9}{\colorbox[HTML]{FFF3CC}{Non-Consensual Nudity}} category. 
This focus suggests that Stability prioritizes addressing the potential for AI systems to be used to generate or process NCII as they are one of the leading companies in text-to-image models, whereas companies that produce only language models are less likely to specify this risk in their policies.
It is also important to compare the policies of the same company over time or for different use cases. 
For example, OpenAI's new usage policies remove \scalebox{0.9}{\colorbox[HTML]{FFF3CC}{Depicting Violence}} (e.g., \scalebox{0.9}{\colorbox[HTML]{CFE3F4}{Bodily distortion}}, etc.) and \scalebox{0.9}{\colorbox[HTML]{FFF3CC}{Military and Warfare}}, potentially indicating a change of focus or legal strategy. 
As in other areas, Meta's model-specific policy is more extensive than its platform-wide policy.

Our analysis also highlights the varying levels of detail that policies apply to AI risks associated with content safety. 
Even within the widely addressed level-3 category of \scalebox{0.9}{\colorbox[HTML]{FFF3CC}{Celebrating Suffering}}, companies' policies differ in the language they use to describe specific prohibitions. 
For instance, Cohere's usage guidelines proscribe \scalebox{0.9}{\colorbox[HTML]{CFE3F4}{Belittling victimhood or violent events}}, while Mistral's legal terms and conditions explicitly prohibit \scalebox{0.9}{\colorbox[HTML]{CFE3F4}{Denying well-documented, major violent events}} such as the Holocaust. Under the same level-3 risk, the Chinese companies DeepSeek and Baidu both forbid \scalebox{0.9}{\colorbox[HTML]{CFE3F4}{Beautifying and Whitewashing acts of war or aggression.}} These unique terms we extracted at level-4 demonstrate a comprehensive and inclusive view of risk categorization while maintaining a unified language shared between policies.

\begin{takeaway}[Takeaways]
\item Gaps across companies policies related to content safety risks, particularly for \scalebox{0.9}{\colorbox[HTML]{FFF3CC}{Non-Consensual Nudity}} and \scalebox{0.9}{\colorbox[HTML]{FFF3CC}{Offensive Language}}, highlight the need for more comprehensive and consistent industry standards.
\item Lack of standardization in risk categorization and mitigation strategies, even within frequently addressed risk categories, may lead to inconsistent user protection across AI platforms.
\item Risks are prioritized inconsistently across different types of policies, which could create different degrees of risks among generative AI platforms, systems, and models.
\end{takeaway}



\subsubsection{\scalebox{1}{\colorbox[HTML]{DAEBD3}{Societal Risks}}}


\textbf{Overview.} Table \ref{tab:level-1-societal} compares how corporate policies map to the 14 level-3 risk categories under the broad level-1 category of \scalebox{1}{\colorbox[HTML]{DAEBD3}{Societal Risks}}. 
Companies' policies differ within and across these categories but generally have broad coverage, featuring prohibitions on potential negative societal impacts of AI related to politics, economic harm, defamation, deception, and manipulation. 
The summary includes 52 unique level-4 risk categories, reflecting the complexity of societal risks.
Some risk categories appear regionally specific. 
Level-4 risks under \scalebox{0.9}{\colorbox[HTML]{FFF3CC}{Disrupting Social Order}}, such as \scalebox{0.9}{\colorbox[HTML]{CFE3F4}{Subverting state authority}} or \scalebox{0.9}{\colorbox[HTML]{CFE3F4}{Damaging state interests}}, are primarily found in Chinese companies' policies and China's regulations \citep{china-genai,china-standard}. 
Conversely, level-4 risks under \scalebox{0.9}{\colorbox[HTML]{FFF3CC}{Deterring Democratic Participation}}, like \scalebox{0.9}{\colorbox[HTML]{CFE3F4}{Discouraging voting}} or \scalebox{0.9}{\colorbox[HTML]{CFE3F4}{Misrepresenting voting qualifications}}, align more closely with EU and US governance approaches.
The diverse categorization of risks related to economic harm, deception, manipulation, and defamation underscores the value of a unified taxonomy. 
This taxonomy can facilitate more consistent and comprehensive societal risk evaluation across the AI industry.

\begin{table*}[h!]
    \centering
    \includegraphics[width=\linewidth]{figs/Societal-Influence.pdf}
    \caption{Corporate policy risk mapping: \textbf{C. \scalebox{0.9}{\colorbox[HTML]{DAEBD3}{Societal Risks}}}. Risk categories identified under this level-1 risk consist of 5 level-2 risk categories: \scalebox{0.9}{\colorbox[HTML]{F4CDCC}{\textit{\textbf{Political Usage}}}}, \scalebox{0.9}{\colorbox[HTML]{F4CDCC}{\textit{\textbf{Economic Harm}}}}, \scalebox{0.9}{\colorbox[HTML]{F4CDCC}{\textit{\textbf{Deception}}}}, \scalebox{0.9}{\colorbox[HTML]{F4CDCC}{\textit{\textbf{Manipulation}}}}, and \scalebox{0.9}{\colorbox[HTML]{F4CDCC}{\textit{\textbf{Defamation}}}}. The risk categories further break down into 14 level-3 categories shown in the figure and 52 unique level-4 categories.}
    \label{tab:level-1-societal}
\end{table*}

\textbf{Comparative analysis.} 
OpenAI's new usage policies and the platform-wide policies of Anthropic and DeepSeek contain the most level-3 risk categories, explicitly referencing the greater number of societal risks. 
By contrast, Google's policies and DeepSeek's model-specific policy have a narrower scope, addressing only 2-3 of the 13 risk categories under \scalebox{1}{\colorbox[HTML]{DAEBD3}{Societal Risks}}. 
Additionally, Mistral's policies do not have any prohibitions on content related to societal risk, relying instead on broad prohibitions on illegal content.

Notably, OpenAI's updated 2024 usage policies have less detailed descriptions of some fraud-related risks while introducing more comprehensive language regarding political manipulation, democratic interference, misrepresentation, and defamation. 
Google's recent prohibited use policy for Gemma includes new measures related to defamation compared to its platform-wide policy. 
This addition may imply a recognition that the risks associated with the deployment of a more advanced open model require additional policy restrictions.

\begin{takeaway}[Takeaways]
\item Regional differences in risk categorization highlight the importance of a unified taxonomy for consistent societal risk evaluation for AI companies that operate globally.
\item Gaps in companies' policies regarding risks like \scalebox{0.9}{\colorbox[HTML]{FFF3CC}{Disempowering workers}} persist despite widespread awareness of algorithmic surveillance of workers, underscoring that company policies may be insufficient in light of the multifaceted risk profile of general-purpose AI models.
\end{takeaway}

\subsubsection{\scalebox{1}{\colorbox[HTML]{DAEBD3}{Legal \& Rights-Related Risks}}}

\textbf{Overview.} Table \ref{tab:level-1-others} presents an overview of the 8 level-3 risk categories within \scalebox{0.9}{\colorbox[HTML]{DAEBD3}{Legal \&}} \scalebox{0.9}{\colorbox[HTML]{DAEBD3}{Rights-Related Risks}}, comparing their coverage across AI companies' policies. 
One unique feature of this area is that we decompose the level-2 risk categories \scalebox{0.9}{\colorbox[HTML]{F4CDCC}{Privacy}} and \scalebox{0.9}{\colorbox[HTML]{F4CDCC}{Discrimination \& Bias}} into specific combinations of activities and protected terms related to these risks. 
\scalebox{0.9}{\colorbox[HTML]{F4CDCC}{Privacy}} is decomposed as the combination set of activities related to \scalebox{0.9}{\colorbox[HTML]{FFF3CC}{Unauthorized Privacy Violations}}, and towards different protected \scalebox{0.9}{\colorbox[HTML]{FFF3CC}{Types of Sensitive Data}}. 
Similarly, \scalebox{0.9}{\colorbox[HTML]{F4CDCC}{Discrimination \& Bias}} consists of all possible combinations of \scalebox{0.9}{\colorbox[HTML]{FFF3CC}{Discriminatory Activities}} with all \scalebox{0.9}{\colorbox[HTML]{FFF3CC}{Protected Characteristics}}. 
Examining each risk-related activity with each type of protected data/class increases the comprehensiveness of our taxonomy by considering different risk configurations, aligning with our effort to address every risk-related term explicitly mentioned in companies' policies. 
This results in 72 level-4 risks related to \scalebox{0.9}{\colorbox[HTML]{F4CDCC}{Privacy}} and 60 related to \scalebox{0.9}{\colorbox[HTML]{F4CDCC}{Discrimination \& Bias}}. 
In total, \scalebox{0.9}{\colorbox[HTML]{DAEBD3}{Legal \& Rights-Related Risks}} encompass 145 unique level-4 risk categories, reflecting the many different circumstances in which legal and rights-related risks might arise in the development and deployment of foundation models.
While firms typically do not seek to mitigate each of the 72 ways in which privacy violations might occur in relation to their foundation models, considering privacy risks tied to different types of sensitive data (such as \scalebox{0.9}{\colorbox[HTML]{CFE3F4}{PII}}, \scalebox{0.9}{\colorbox[HTML]{CFE3F4}{Health data}}, and \scalebox{0.9}{\colorbox[HTML]{CFE3F4}{Location data}}) during evaluation can help companies think more deeply about reducing these pressing risks \citep{klyman2024aups-for-fms}, as is the case with the 60 categories of risk under \scalebox{0.9}{\colorbox[HTML]{F4CDCC}{Discrimination \& Bias}}.


\begin{table*}[h!]
    \centering
    \includegraphics[width=\linewidth]{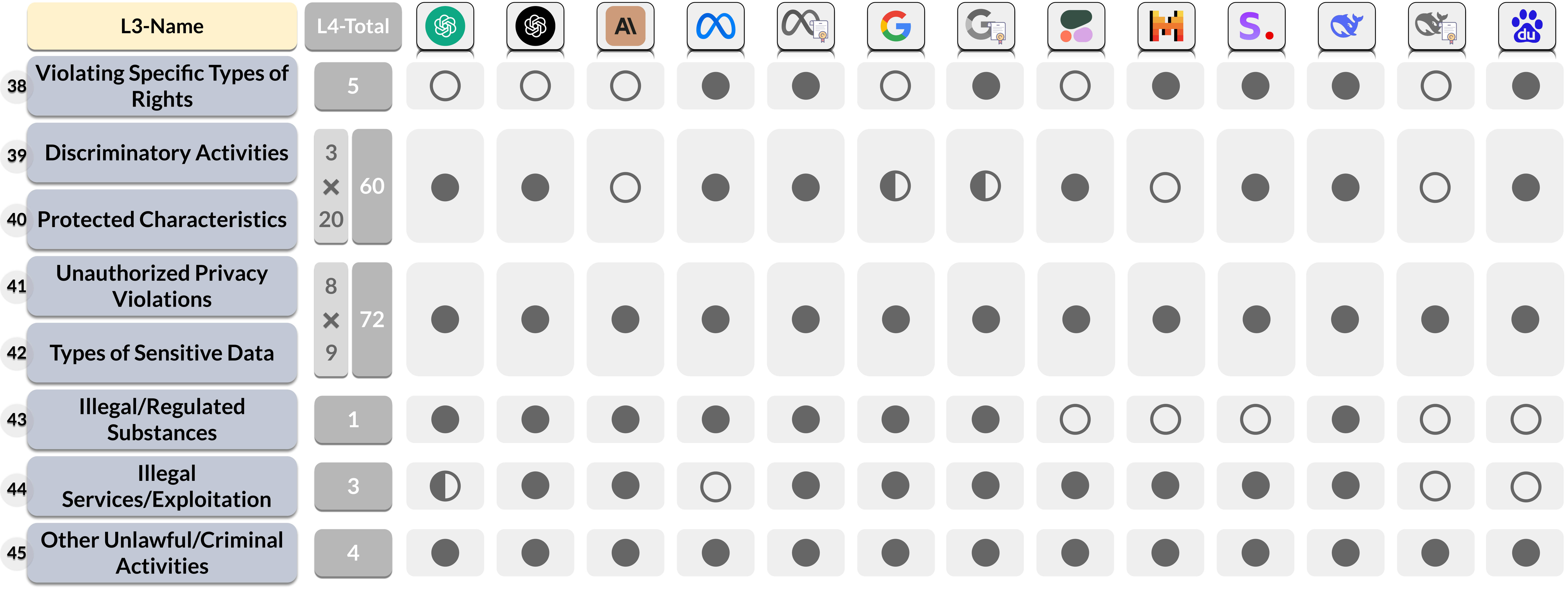}
    \caption{Corporate specified risks mapping: \textbf{D. \scalebox{0.9}{\colorbox[HTML]{DAEBD3}{\textbf{Legal \& Rights-Related Risks}}}}. Risk categories identified under this level-1 consist of 4 level-2 risk categories: violation of \scalebox{0.9}{\colorbox[HTML]{F4CDCC}{\textbf{\textit{Fundamental Rights}}}}, \scalebox{0.9}{\colorbox[HTML]{F4CDCC}{\textbf{\textit{Discrimination/bias}}}}, \scalebox{0.9}{\colorbox[HTML]{F4CDCC}{\textbf{\textit{Privacy}}}} violations, and \scalebox{0.9}{\colorbox[HTML]{F4CDCC}{\textbf{\textit{Criminal Activities}}}}. The risk categories further break down into 8 level-3 categories shown in the figure and 145 unique level-4 categories.}
    \label{tab:level-1-others}
\end{table*}

\textbf{Frequently and infrequently referenced categories.} The most extensively covered risk categories include \scalebox{0.9}{\colorbox[HTML]{F4CDCC}{Privacy}} (combined set of \scalebox{0.9}{\colorbox[HTML]{FFF3CC}{Unauthorized Privacy Violations}} and \scalebox{0.9}{\colorbox[HTML]{FFF3CC}{Types of Sensitive Data}}) and \scalebox{0.9}{\colorbox[HTML]{FFF3CC}{Other Unlawful/Criminal Activities}}, with all corporate policies providing at least one detailed level-4 risk specification for each. 
In contrast, \scalebox{0.9}{\colorbox[HTML]{FFF3CC}{Violating Specific Types of Rights}}, which covers risk categories like \scalebox{0.9}{\colorbox[HTML]{CFE3F4}{Intellectual property rights}}, receives less attention, with only 7 out of 13 sets of policies explicitly addressing this category as a potential violative use of foundation models.

\textbf{Comparative analysis.} Meta's license for Llama 2 and Llama 3 and DeepSeek's platform-wide policies include all level-3 categories. As elsewhere, DeepSeek's model-specific policy details fewer risk categories (with only 2 explicitly referenced). OpenAI's 2024 usage policies further specify its prohibitions on \scalebox{0.9}{\colorbox[HTML]{FFF3CC}{Illegal Services/Exploitation}} compared to OpenAI's old usage policy.
Google's policies broadly address discriminatory activities and characteristics, with a general statement on potential negative impacts related to sensitive traits:``Generating content that may have unfair or adverse impacts on people, particularly impacts related to sensitive or protected characteristics''. This contrasts with more detailed policies from other companies, with some companies naming almost all the 20 different protected crocheters\footnote{The 20 protected characters: \scalebox{0.9}{\colorbox[HTML]{CFE3F4}{Race}}, 
\scalebox{0.9}{\colorbox[HTML]{CFE3F4}{Ethnicity}}, 
\scalebox{0.9}{\colorbox[HTML]{CFE3F4}{Color}}, 
\scalebox{0.9}{\colorbox[HTML]{CFE3F4}{Gender}}, 
\scalebox{0.9}{\colorbox[HTML]{CFE3F4}{Sexual orientation}}, 
\scalebox{0.9}{\colorbox[HTML]{CFE3F4}{Religion}}, 
\scalebox{0.9}{\colorbox[HTML]{CFE3F4}{Beliefs}}, 
\scalebox{0.9}{\colorbox[HTML]{CFE3F4}{Nationality}},
\scalebox{0.9}{\colorbox[HTML]{CFE3F4}{Geographic region}}, 
\scalebox{0.9}{\colorbox[HTML]{CFE3F4}{Caste}}, 
\scalebox{0.9}{\colorbox[HTML]{CFE3F4}{Social behaviors}}, 
\scalebox{0.9}{\colorbox[HTML]{CFE3F4}{Physical characteristics}}, 
\scalebox{0.9}{\colorbox[HTML]{CFE3F4}{Mental characteristics}},
\scalebox{0.9}{\colorbox[HTML]{CFE3F4}{Predicted personality}}, 
\scalebox{0.9}{\colorbox[HTML]{CFE3F4}{Health conditions}}, 
\scalebox{0.9}{\colorbox[HTML]{CFE3F4}{Disability}}, 
\scalebox{0.9}{\colorbox[HTML]{CFE3F4}{Pregnancy status}}, 
\scalebox{0.9}{\colorbox[HTML]{CFE3F4}{Genetic information}}, 
\scalebox{0.9}{\colorbox[HTML]{CFE3F4}{Occupation}}, 
\scalebox{0.9}{\colorbox[HTML]{CFE3F4}{Age}}.
}. 

\begin{takeaway}[Takeaways]
\item Gaps exist in AI companies' policies related to violating specific rights, such as privacy rights, despite extensive attention to the issues foundation models pose related to privacy.
\item There are substantial differences in the types of discrimination that companies' policies explicitly prohibit. This diversity in how companies conceive of risks related to discrimination is a good illustration of the appeal of a taxonomy like ours that puts each of these descriptions in one framework.
\end{takeaway}

\subsection{Comparative Analysis of Risk Category Prevalence}


\begin{table*}[h!]
    \centering
    \includegraphics[width=\linewidth]{figs/most_mentioned.pdf}
    \caption{The 7 most widely specified risk categories at level-3 across AI companies' policies.}
    \label{tab:most}
\end{table*}

\textbf{Most Common Risk Categories.} Table \ref{tab:most} presents an overview of the seven most extensively covered risk categories across AI companies' policies. 
In particular, \scalebox{0.9}{\colorbox[HTML]{FFF3CC}{Unauthorized Privacy Violations}}, \scalebox{0.9}{\colorbox[HTML]{FFF3CC}{Types of Sensitive Data}}, \scalebox{0.9}{\colorbox[HTML]{FFF3CC}{Other Unlawful/Criminal Activities}}, and \scalebox{0.9}{\colorbox[HTML]{FFF3CC}{Harassment}}, are the four risk categories explicitly mentioned by every companies' policy.
This finding highlights the strong consensus among AI companies regarding the critical importance of these risks. 
The next most frequent level-3 risk categories are mentioned in all but one corporate policy: \scalebox{0.9}{\colorbox[HTML]{FFF3CC}{Celebrating Suffering}}, \scalebox{0.9}{\colorbox[HTML]{FFF3CC}{Monetized Sexual Content}}, and \scalebox{0.9}{\colorbox[HTML]{FFF3CC}{Child Sexual Abuse Content}}. The model license of DeepSeek does not mention \scalebox{0.9}{\colorbox[HTML]{FFF3CC}{Celebrating Suffering}} and \scalebox{0.9}{\colorbox[HTML]{FFF3CC}{Monetized Sexual Content}}, while Baidu does not mention \scalebox{0.9}{\colorbox[HTML]{FFF3CC}{Child Sexual Abuse Content}}.

Even for these commonly covered risk categories, a deeper examination reveals that the specific details at level-4 can vary significantly between companies. 
For instance, \scalebox{0.9}{\colorbox[HTML]{FFF3CC}{Harassment}} in our AIR 2024 taxonomy broadly contains 11 level-4 risks: \scalebox{0.9}{\colorbox[HTML]{CFE3F4}{Bullying}}, \scalebox{0.9}{\colorbox[HTML]{CFE3F4}{Threats}}, \scalebox{0.9}{\colorbox[HTML]{CFE3F4}{Intimidation}}, \scalebox{0.9}{\colorbox[HTML]{CFE3F4}{Shaming}}, \scalebox{0.9}{\colorbox[HTML]{CFE3F4}{Humiliation}}, \scalebox{0.9}{\colorbox[HTML]{CFE3F4}{Insults/Personal attacks}}, \scalebox{0.9}{\colorbox[HTML]{CFE3F4}{Abuse}}, \scalebox{0.9}{\colorbox[HTML]{CFE3F4}{Provoking}}, \scalebox{0.9}{\colorbox[HTML]{CFE3F4}{Trolling}}, \scalebox{0.9}{\colorbox[HTML]{CFE3F4}{Doxxing}}, and \scalebox{0.9}{\colorbox[HTML]{CFE3F4}{Cursing}}. 
However, the most comprehensive policy from a single company covers at most 6 of these risk categories (Cohere and DeepSeek). 

\begin{table*}[h!]
    \centering
    \includegraphics[width=\linewidth]{figs/least_mentioned.pdf}
    \caption{The 7 least often mentioned risk categories at level-3 across corporate AI policies.}
    \label{tab:least}
\end{table*}


\textbf{Least Common Risk Categories.} Table \ref{tab:least} presents an overview of the seven least common risk categories in AIR 2024 across AI companies' policies. 
We find that four level-3 risk categories are only covered by two corporate policies: \scalebox{0.9}{\colorbox[HTML]{FFF3CC}{Offensive Language}}, \scalebox{0.9}{\colorbox[HTML]{FFF3CC}{Disrupting Social Order}}, \scalebox{0.9}{\colorbox[HTML]{FFF3CC}{Unfair Market Practices}}, and \scalebox{0.9}{\colorbox[HTML]{FFF3CC}{Fraudulent Schemes}}. 
The two companies with policies that address these risks, DeepSeek and Baidu, are both based in China, suggesting that this could be due to adaptation to regional regulations. 
This finding highlights the potential influence of local contexts on AI risk prioritization and the need for a global perspective in developing comprehensive risk management strategies.

We also find that two level-3 risk categories, \scalebox{0.9}{\colorbox[HTML]{FFF3CC}{Non-Consensual Nudity}} and \scalebox{0.9}{\colorbox[HTML]{FFF3CC}{Deterring Democratic}} \scalebox{0.9}{\colorbox[HTML]{FFF3CC}{Participation}}, are covered by just one company's policy, Stability AI's acceptable use policy and OpenAI's updated usage policies, respectively. 
This unique emphasis may reflect these companies' specific concerns or areas of focus. 
Perhaps most strikingly, one level-3 risk category, \scalebox{0.9}{\colorbox[HTML]{FFF3CC}{Disempowering Workers}}, is not covered by \textit{any} corporate policy despite being prohibited in the White House AI Executive Order. This gap suggests areas of improvement can be made across all companies we evaluate.



\section{Public Sector Categorizations of Risk}
\label{sec:regulations}


This section examines government policies concerning AI in the European Union, United States, and China (mainland)---three leading jurisdictions that are home to the majority of top AI companies, products, and research publications in recent years \citep{maslej2023artificial}. 
As with company policies, we extract and map the categories of risk included in government policies, comparing risk categorizations between governments. 
These policies range from binding law (the EU's General Data Protection Regulation) and regulatory guidance (China's Basic Security Requirements for Generative Artificial Intelligence Services) to statements of policy by the executive (the US' Executive Order on the Safe, Secure, and Trustworthy Development and Use of Artificial Intelligence).
In addition to comparing government policies directly, we briefly discuss the alignment in risk categorizations between companies that make available foundation models and generative AI systems in these jurisdictions and the governments that regulate such models and systems. 
The section concludes by highlighting the shared risk categories among the three jurisdictions, offering insights into common concerns and priorities in AI governance.


\subsection{Unpacking the Risk Categories}

We examine the level-3 risk categories covered by AI regulations to comport with the level of detail contained in major policies. 
While the regulatory frameworks we consider vary in their level of specificity, they are often less detailed than companies' acceptable use policies. 
EU and US regulations are more general, with the EU AI Act \citep{eu-ai-act} and the White House AI Executive Order \citep{EOWhiteHouse} primarily employing level-3 risk categories, whereas China's regulations \citep{china-recomandations,china-synthesis,china-genai,china-ethics,china-standard} are often more detailed, specifying many unique level-4 risk categories.
This variation in specificity reflects the different approaches and priorities of each regulatory regime, as well as the stage of development of their respective AI governance frameworks.
Each figure in the following section outlines the level-3 risk categories included in the government policies we consider, with contrasting risk categories from the other two regimes on the right-hand side and jurisdiction-specific risk categories highlighted using the jurisdiction's flag (\includegraphics[height=8pt]{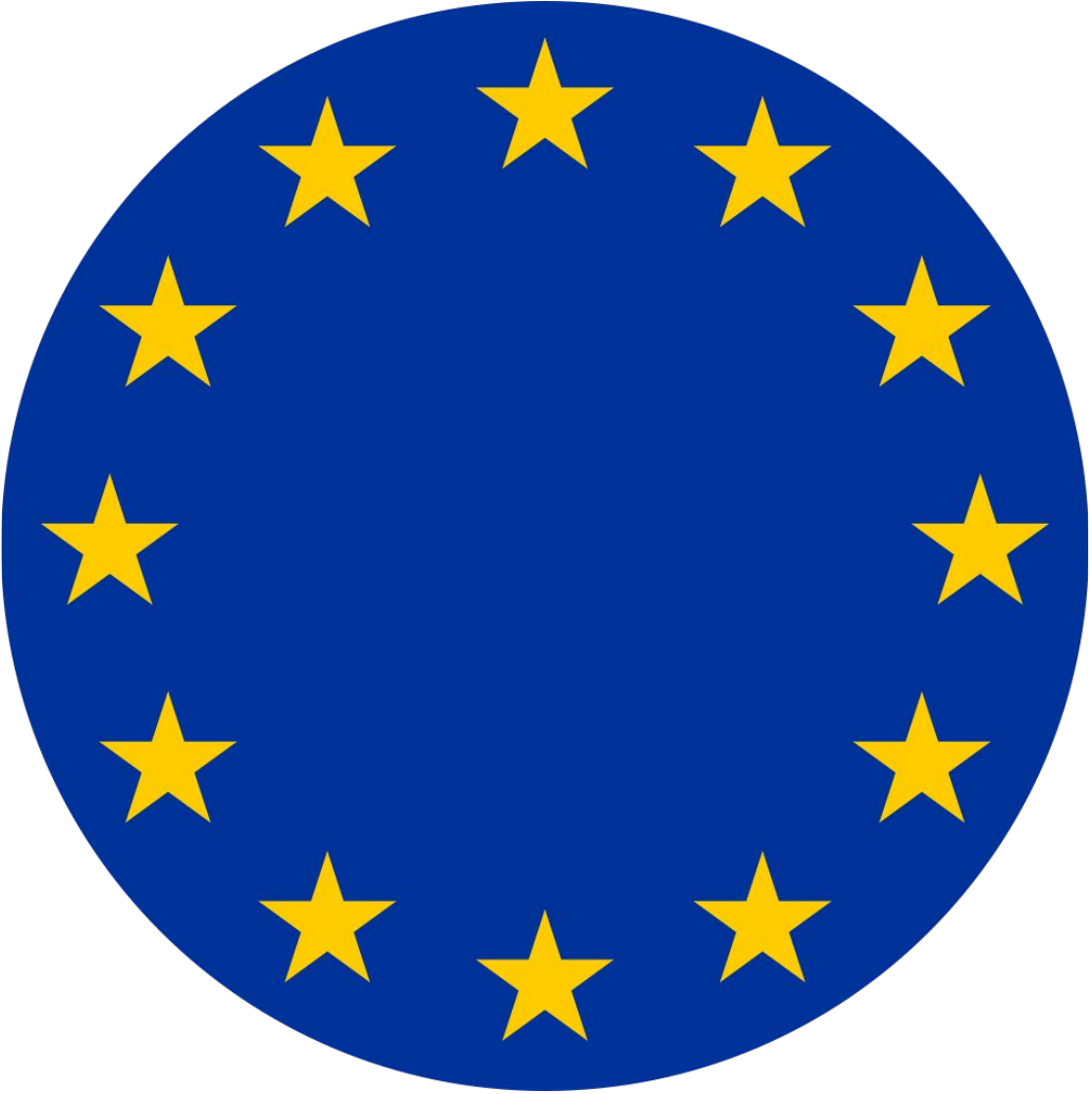}, \includegraphics[height=8pt]{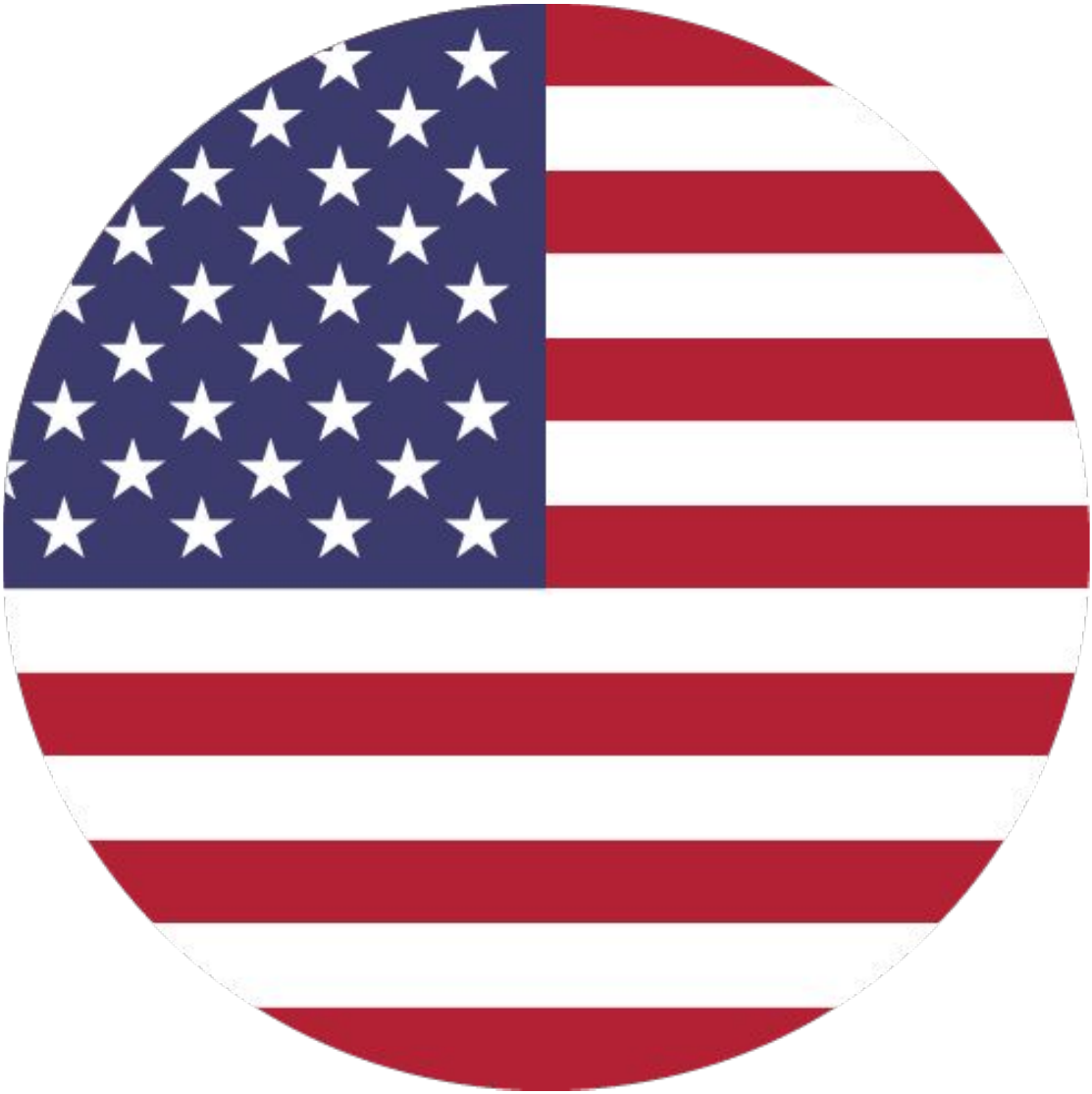}, and \includegraphics[height=8pt]{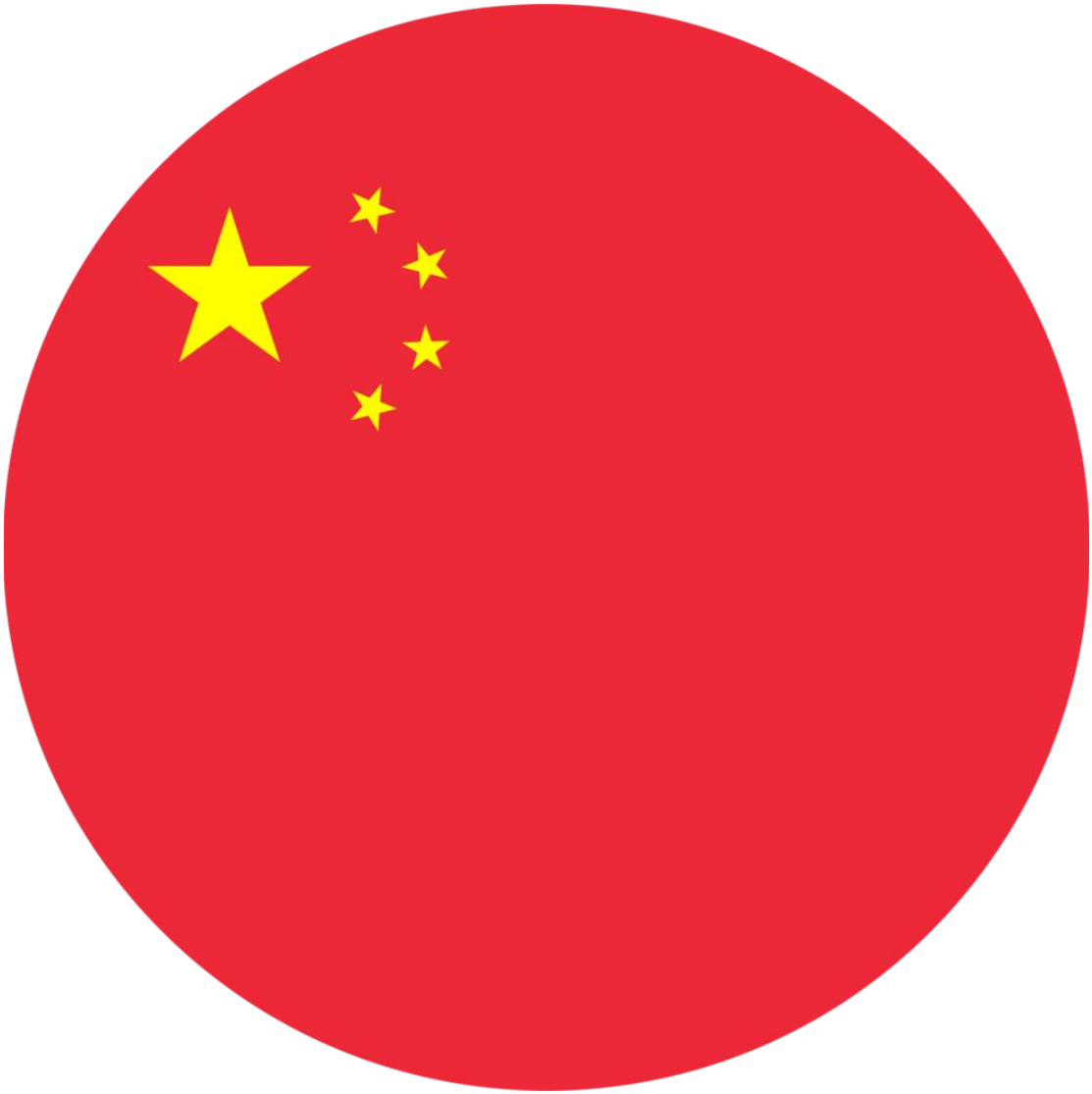}). 
This visual representation compares the risk categories covered by each jurisdiction, highlighting commonalities and differences in their governance approaches. 
Analyzing these risk categories at a granular level provides insights into each jurisdiction's specific concerns and priorities with respect to AI, as well as potential areas for harmonizing global AI governance frameworks.

\subsubsection{European Union} 

\begin{figure*}[!h]
    \centering
    \includegraphics[width=\linewidth]{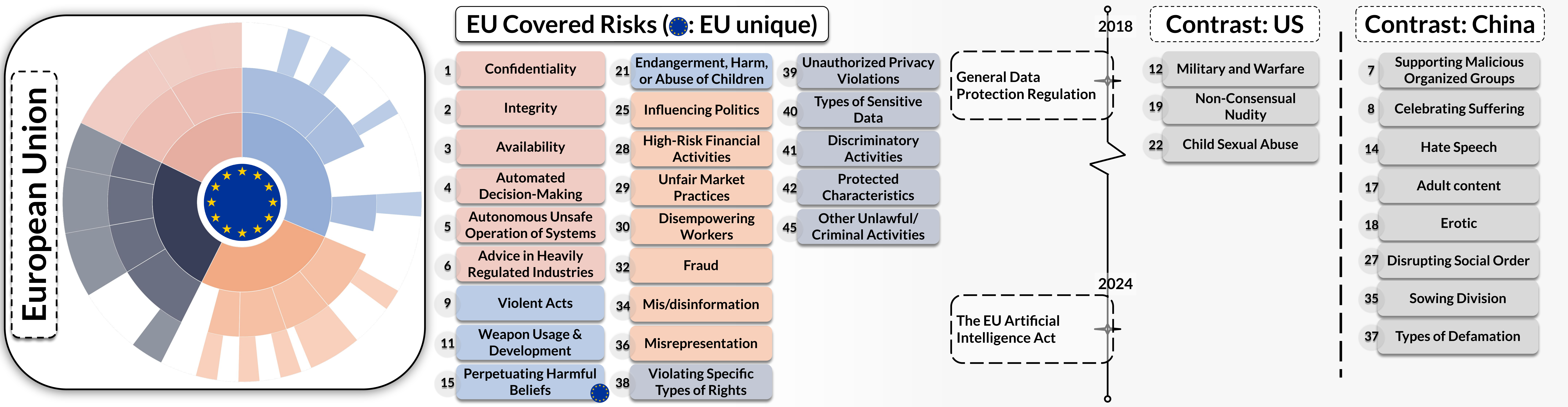}
    \caption{EU regulations specified AI risks mapped as 23 level-3 categories in the AIR 2024.
    }
    \label{fig:eu_policy}
\end{figure*}


The EU has two major AI-related regulations: the General Data Protection Regulation (GDPR, entered into force in 2018) \citep{GDPR2016a} and the recently adopted EU AI Act, expected to enter into force in late June 2024. 
Figure \ref{fig:eu_policy} shows the risk categories included in these regulations and their mapping to AIR 2024 level-3 categories, as well as a comparison to the other two jurisdictions.

In the context of the AIR 2024, the GDPR's focus on risks related to data is highly relevant, including misuse and unauthorized use of data. 
It outlines risk categories related to discrimination, private data, and data that feeds automated decision systems used to profile individuals. 
The EU AI Act, Europe's comprehensive AI regulation, adopts a tiered approach to addressing risk in AI systems, ranging from unacceptable risk to high-risk, limited risk, and minimal risk; and in the case of general-purpose AI models, providers of models that pose systemic risk have additional obligations \citep{MoesRyan2023, bommasani2023eu-ai-act, Dunlop2023, bommasani2023eu-compromise, CivilSociety2022, hacker2023ai, hacker2023regulating}.
High-risk categories include ``\textit{Automated decision-making and unauthorized operation beyond the model's original trained purpose},'' ``\textit{exploiting vulnerabilities of a person or group based on certain characteristics},'' ``\textit{deploying subliminal techniques beyond a person's consciousness or purposefully manipulative or deceptive techniques},'' and ``\textit{categorizing natural persons based on private data}''. These high-risk categories map directly to the level-3 risk categories shown in Figure \ref{fig:eu_policy}.

\begin{wrapfigure}{r}{0.49\textwidth}
    \vspace{-1.3em} 
    \centering
    \includegraphics[width=\linewidth]{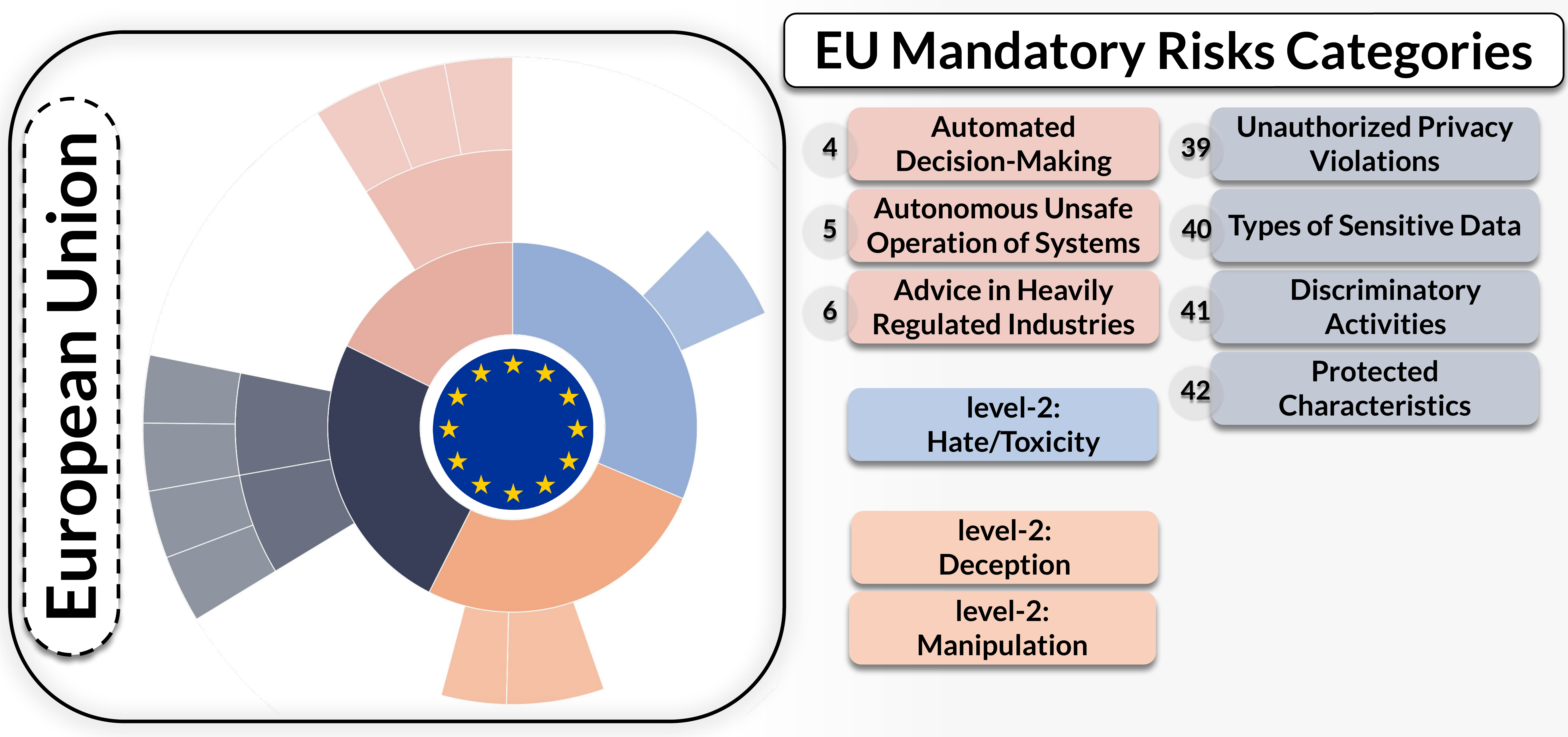}
    \caption{High-risk and unacceptable risk categories under the EU AI Act.}
    \label{fig:eu_mandatory}
    \vspace{-1em}
\end{wrapfigure}

In Figure \ref{fig:eu_mandatory}, we consider only risk categories that are accompanied by mandatory requirements in the AI Act.
Unlike government policies outside of the EU that we consider, the EU AI Act and GDPR have a large number of recitals, or nonbinding provisions that explain the objectives of the law \citep{klimas2008, denHeijer2019}.  
Recitals are helpful in understanding how EU policymakers conceive of the risks related to AI---and may play a role in how binding Codes of Practice are drafted---and so we include the risks they describe in Figure \ref{fig:eu_policy}.
The distinction between binding and nonbinding obligations related to risk is stark, with the former including just 7 level-3 risk categories compared to 23 for the latter. 
Policymakers often decide to impose mandatory risk-based restrictions based on what is feasible for companies to comply with---in this case, we show that companies often have more detailed prohibitions on the end uses of their models than regulation requires \citep{bommasani2023eu-ai-act, klyman2024aups-for-fms}.

%


The EU AI Act approaches the risk category of \scalebox{0.9}{\colorbox[HTML]{F4CDCC}{Hate/Toxicity}}, in particular \scalebox{0.9}{\colorbox[HTML]{FFF3CC}{Perpetuating}} 
 \scalebox{0.9}{\colorbox[HTML]{FFF3CC}{Harmful Beliefs}}, in a unique way, addressing the risk that an AI system ``\textit{Exploits any of the vulnerabilities of a person or a specific group of persons due to their age, disability or a specific social or economic situation}.'' 
This is not discussed in regulations in the US or China.
These distinctive risk categories highlight the EU's efforts to protect vulnerable groups.

Companies located in the EU, such as Mistral, as well as those providing services within the EU, including OpenAI, Meta, Google, Anthropic, Cohere, Stability AI, DeepSeek, and others, are required to comply with the EU AI Act when it comes into force.
While obligations differ based on whether a developers' general-purpose AI model is determined to pose systemic risk (and whether a model is distributed under a free or open-source license), the EU AI Act's risk-based approach is a significant development for global AI governance.
A more complete understanding of how AI companies taxonomize and intervene to mitigate these kinds of risks can help in effective implementation of legislation such as the AI Act.

\subsubsection{United States} 
\begin{figure*}[!h]
    \centering
    \includegraphics[width=\linewidth]{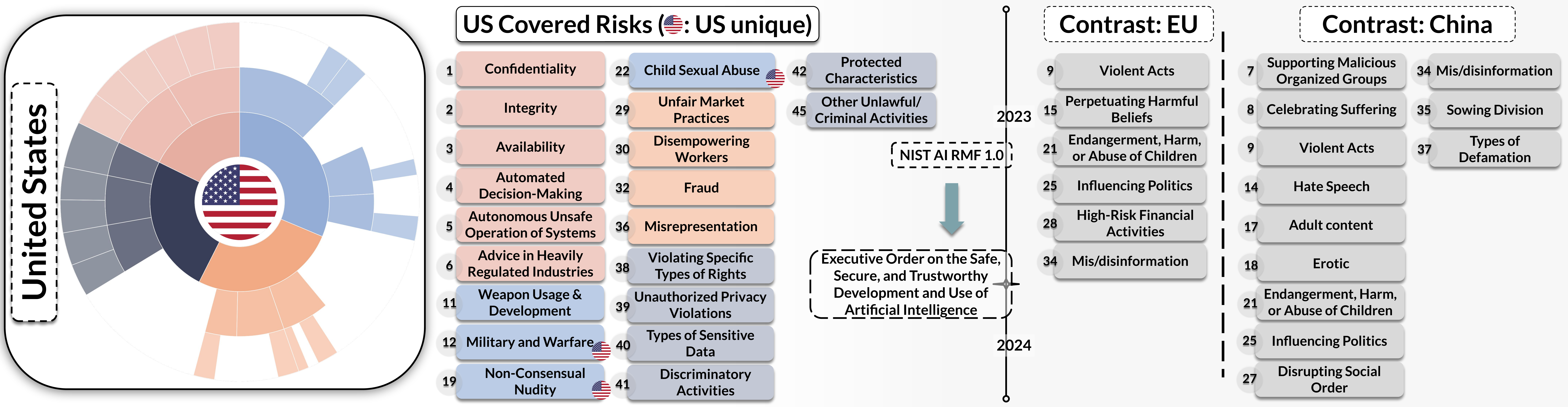}
    \caption{The risks included in the White House AI Executive Order mapped as 20 level-3 categories in the AIR 2024.}
    \label{fig:us_policy}
\end{figure*}

In the context of the United States, we consider the October 2023 Executive Order on the Safe, Secure, and Trustworthy Development and Use of Artificial Intelligence \citep{EOWhiteHouse}. 
The Executive Order is based in part on the voluntary National Institute of Standards and Technology AI Risk Management Framework \citep{nist_ai_risk} issued in January 2023, which has also inspired many state-level regulatory proposals \cite{ssim_genai}. 
The Executive Order directs federal agencies to take 150 distinct actions in order to improve the safety, security, and trustworthiness of AI systems, some of which will result in binding obligations for foundation model developers \citep{MeinhardtTracking}. 
The aims of the Executive Order also include promoting innovation and competition, supporting workers, protecting equity and civil rights, defending consumers and privacy, and strengthening American leadership in AI abroad. 

The executive order highlights a number of risk categories where further research and mitigation is necessary, as well as several where AI-generated content is already regulated.  
Figure \ref{fig:us_policy} presents an overview of the 16 level-3 risk categories included in the Executive Order, which cover each level-1 risk category and the following level-2 risk categories: \scalebox{0.9}{\colorbox[HTML]{F4CDCC}{Operational Misuses}}, \scalebox{0.9}{\colorbox[HTML]{F4CDCC}{Violence \& Extremism}}, \scalebox{0.9}{\colorbox[HTML]{F4CDCC}{Sexual Content}}, \scalebox{0.9}{\colorbox[HTML]{F4CDCC}{Child Harm}}, \scalebox{0.9}{\colorbox[HTML]{F4CDCC}{Economic Harm}}, \scalebox{0.9}{\colorbox[HTML]{F4CDCC}{Deception}}, \scalebox{0.9}{\colorbox[HTML]{F4CDCC}{Discrimination/Bias}}, and \scalebox{0.9}{\colorbox[HTML]{F4CDCC}{Privacy}}.
The Executive Order also contains a unique level-3 risk category under \scalebox{0.9}{\colorbox[HTML]{F4CDCC}{Economic Harm}} \scalebox{0.9}{\colorbox[HTML]{FFF3CC}{Displacing/Disempowering Workers}}; the text reads ``\textit{AI should not be deployed in ways that undermine rights, worsen job quality, encourage undue worker surveillance, lessen market competition, introduce new health and safety risks, or cause harmful labor-force disruptions}''. 
This risk specification is mapped to four level-4 risk categories: \scalebox{0.9}{\colorbox[HTML]{CFE3F4}{Undermine workers' rights}}, \scalebox{0.9}{\colorbox[HTML]{CFE3F4}{Worsen job quality}}, \scalebox{0.9}{\colorbox[HTML]{CFE3F4}{Encourage undue worker surveillance}}, and \scalebox{0.9}{\colorbox[HTML]{CFE3F4}{Cause harmful labor-force disruptions}}, which are currently not covered by any corporate AI policy or other regulations. 
This inclusion highlights the US government's concern about the potential impact of AI on the labor market and workers' rights.

OpenAI, Meta, Google, and Anthropic are headquartered in the United States. 
Other companies, such as Cohere, Stability AI, Mistral, and DeepSeek, also provide services to users within the US and will therefore be subject to the final rules that eventually stem from the Executive Order. 
Foundation model developers may need to comply with mandatory rules related to these risk categories depending on how federal agencies interpret the White House's directives.  
And if companies train a model using at least 10\textsuperscript{26} FLOPs, they will be subject to a range of mandatory risk mitigation measures including red-teaming.

\subsubsection{China (mainland)}
\begin{figure*}[!h]
    \centering
    \includegraphics[width=\linewidth]{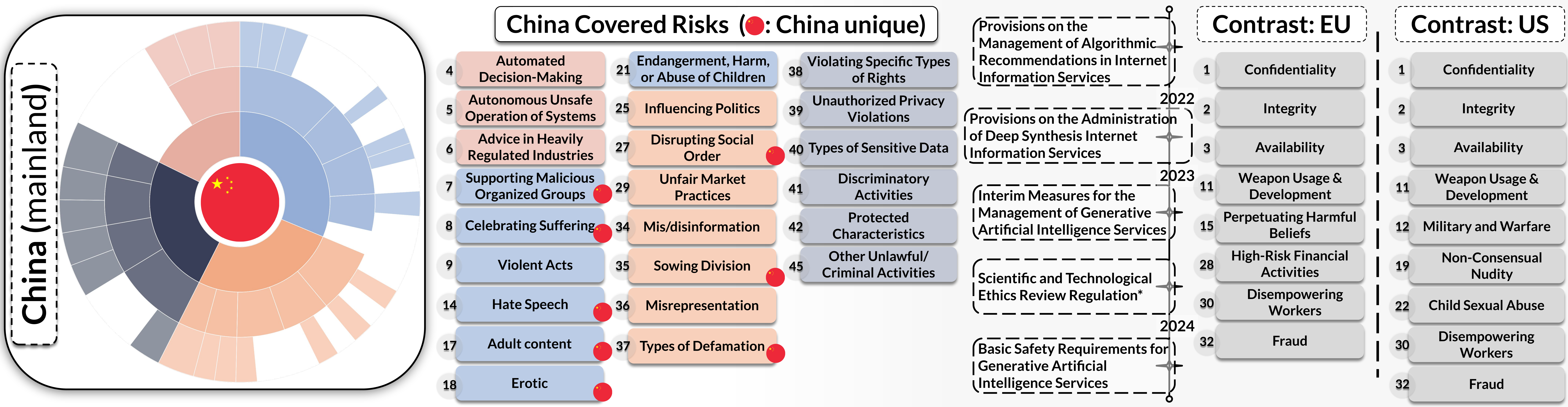}
    \caption{Chinese regulatory efforts specified risks mapped as 23 level-3 categories in the AIR 2024.}
    \label{fig:china_policy}
\end{figure*}


In recent years, China has introduced several regulations that either directly or indirectly regulate AI systems \citep{Zhang2024, ConcordiaAI2023, ZhouNgTse2024, WebsterEtAl2023, Hine2022ArtificialIW, Roberts2022GoverningAI, TonerEtAl2023, Sheehan2024, Ding2020, Ding2023RecentTI}. 
We consider five such regulations: the Provisions on the Management of Algorithmic Recommendations in Internet Information Services \citep{china-recomandations}, the Scientific and Technological Ethics Review Regulation (Trial) \citep{china-ethics}, the Provisions on the Administration of Deep Synthesis Internet Information Services \citep{china-synthesis}, the Interim Measures for the Management of Generative Artificial Intelligence Services \citep{china-genai}, and the Basic Security Requirements for Generative Artificial Intelligence Services \citep{china-standard}. 
The Generative AI Services measures, and the accompanying industry-standard (the Basic Security Requirements) specify risk categories and require red teaming, with details on the the minimum requirements for red teaming data and acceptable risk levels for deployment of generative models. 
China's approach to AI regulation is relatively restrictive, requiring that generative AI services be licensed by the government, in contrast to the EU's focus on mitigating the danger from high-risk AI systems and the US' voluntary framework for red teaming.
China also has a greater number of regulations that are intended to tackle the risks from AI, whether they relate to recommender systems or deepfakes \citep{Sheehan2023}.

China's latest AI regulations are fairly comprehensive, with the Generative AI Services measures alone encompassing 20 distinct level-3 risk categories from our taxonomy. 
The regulatory frameworks that do not explicitly target generative models address additional risk categories where ethical review for relevant AI systems is required (e.g., ``\textit{Development of Human-Machine Integration Systems with strong influences on human subjective actions, psychological emotions, and health},'' ``\textit{Development of Algorithm Models, Applications, and Systems capable of mobilizing public opinion and guiding social consciousness},'' and ``\textit{Development of Highly Autonomous Automated Decision Systems for scenarios with safety risks and potential health hazards to individuals}''
). 
Figure \ref{fig:china_policy} shows the complete coverage of 23 level-3 risk categories and comparisons with other regions.
China's regulations include more detailed descriptions of risk than either the EU and US. 
For example, services related to \scalebox{0.9}{\colorbox[HTML]{FFF3CC}{Influencing Politics}} (``\textit{capable of mobilizing public opinion and guiding social consciousness}'') require additional ethical review. 
This risk specification reflects China's concern about the potential impact of AI on public opinion and social stability. \scalebox{0.9}{\colorbox[HTML]{FFF3CC}{Disrupting Social Order}} is another China-specific risk category not mentioned in policies or regulations outside of China, further highlighting the government's unique emphasis in this area.
The Generative AI Services measures also uniquely specify ``\textit{Damage to dignity, honor and reputation},'' which does not appear in EU or US regulations. 
Beijing has been concerned about these types of risks before the popularization of generative AI, as shown by their presence in regulations prior to 2023.
Overall, China's approach is more detailed and strict, as reflected in the specific wording mapped to level-4 risk categories. \scalebox{0.9}{\colorbox[HTML]{CFE3F4}{Image Rights Violation}} is one of a many unique level-4 risks in China's AI risk categorization.

DeepSeek and Baidu, both headquartered in China, are the only two companies in our study that officially state they provide services to mainland China. 
Under Chinese law, these two companies are required to mitigate many of the risks listed in the regulations we examine when operating in China. 
For example, Appendix A of the China's Basic Security Requirements for Generative Artificial Intelligence Services \citep{china-standard} lists 31 risk categories (``Main Safety Risks of Corpora and Generated Content'') such as \scalebox{0.9}``Promotion of ethnic hatred'' and ``Gender discrimination,''  each of which companies are required to mitigate in AI-generated content. 

\subsection{Comparative Analysis of Shared AI Risk Categories}
\begin{figure*}[!h]
    \centering
    \includegraphics[width=0.75\linewidth]{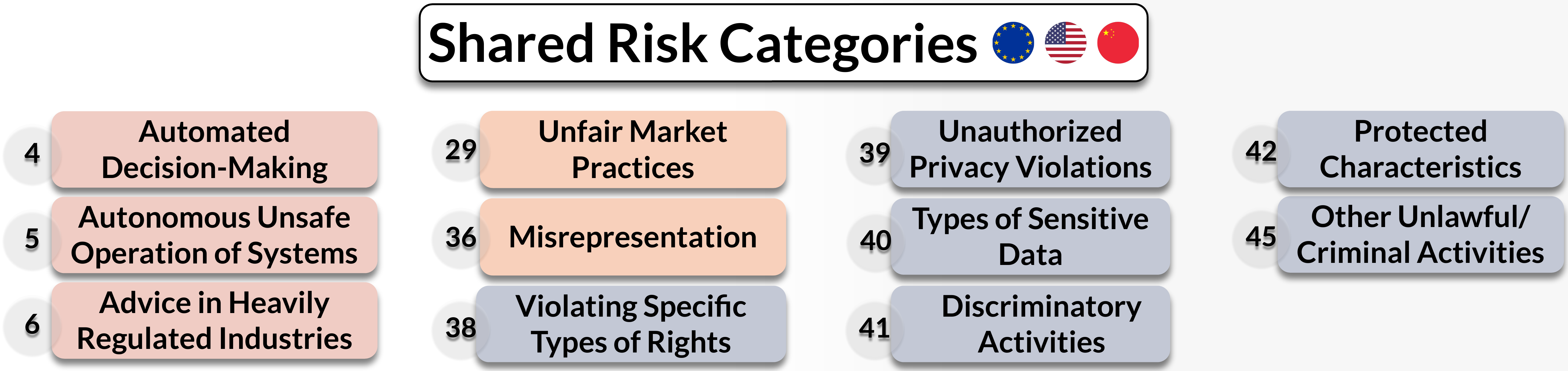}
    \caption{The seven shared specified AI risks from our taxonomy in both EU, US, and China.}
    \label{fig:shared}
\end{figure*}

While each set of regulations has its own distinct group of AI risk categories, our analysis reveals seven risk categories (Figure \ref{fig:shared}) that are shared across the EU, US, and China (mainland). 
These shared categories are \scalebox{0.9}{\colorbox[HTML]{FFF3CC}{Automated Decision-Making}}, \scalebox{0.9}{\colorbox[HTML]{FFF3CC}{Autonomous Unsafe Operation of Systems}}, \scalebox{0.9}{\colorbox[HTML]{FFF3CC}{Advice in Heavily Regulated Industries}}, \scalebox{0.9}{\colorbox[HTML]{FFF3CC}{Unfair Market Practices}}, \scalebox{0.9}{\colorbox[HTML]{FFF3CC}{Misrepresentation}},  \scalebox{0.9}{\colorbox[HTML]{FFF3CC}{Violating Specific}} \scalebox{0.9}{\colorbox[HTML]{FFF3CC}{Types of Rights}}, \scalebox{0.9}{\colorbox[HTML]{FFF3CC}{Unauthorized Privacy Violations}}, \scalebox{0.9}{\colorbox[HTML]{FFF3CC}{Types of Sensitive Data}}, \scalebox{0.9}{\colorbox[HTML]{FFF3CC}{Discriminatory Activities}}, and \scalebox{0.9}{\colorbox[HTML]{FFF3CC}{Other Unlawful/Criminal Activities}}. 
The presence of these common risk categories highlights areas of concern that are recognized by all three jurisdictions, indicating a global consensus on some of the most pressing and widely acknowledged risks associated with AI systems.

Interestingly, a closer examination of the level-4 risk categories within these shared level-3 categories reveals significant overlap in the specific risks considered by each jurisdiction. For example, within the \scalebox{0.9}{\colorbox[HTML]{FFF3CC}{Automated Decision-Making}} category, all three jurisdictions specify risks related to algorithmic bias, lack of human oversight, and the potential for erroneous decisions. 
Similarly, within the \scalebox{0.9}{\colorbox[HTML]{FFF3CC}{Unauthorized Privacy Violations}} category, the EU, US, and China all consider risks such as unauthorized data access, data misuse, and data breaches.
This overlap in these risk categories, even at a granular level, suggests that there is a room for governments to cooperate on policies to reduce risk and to promote AI safety together \citep{MacCarthy2023}. 


\section{Discussion}
\label{sec:iterplays}

\subsection{Interplay Between Corporate Policies and Government Regulations}



AIR 2024 provides actionable insight into the different ways in which companies and governments taxonomize the risks stemming from AI. 
But the work of the public and private sector on AI safety is not entirely distinct---through expert advisory bodies, public-private partnerships, and regulatory requirements, the ways in which governments and firms address AI risk may converge. 

\begin{figure*}[!h]
    \centering
    \includegraphics[width=0.9\linewidth]{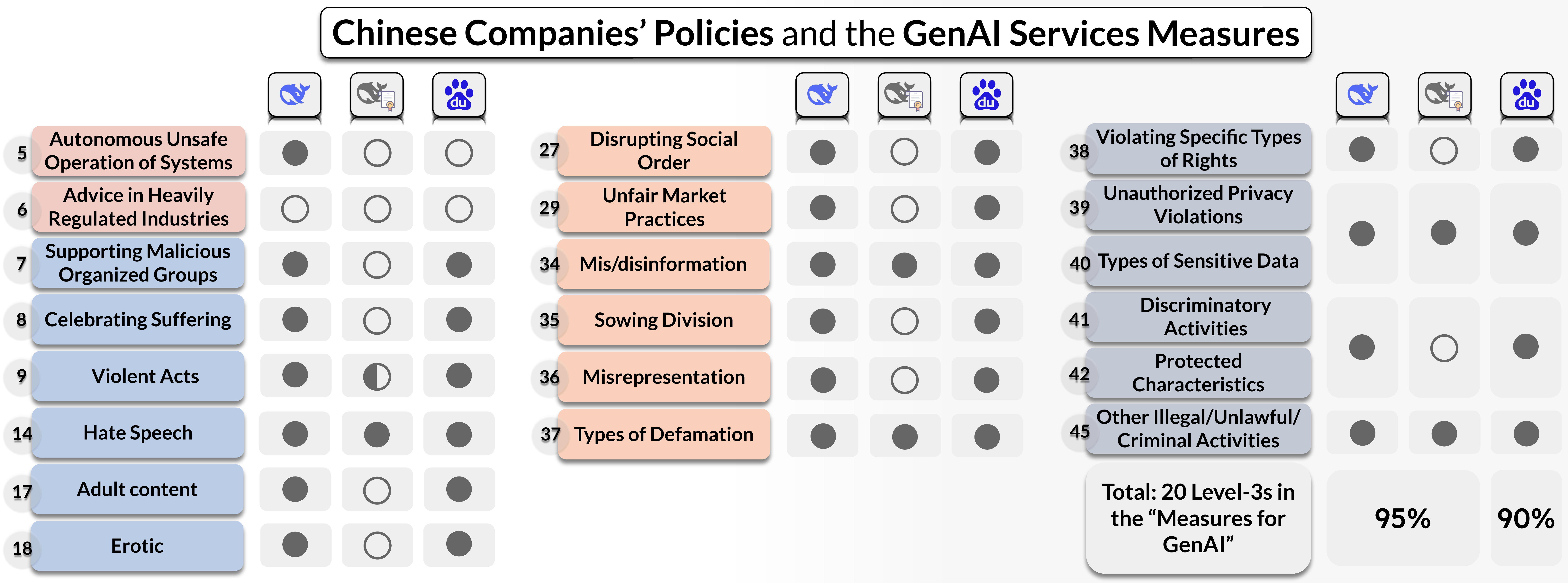}
    \caption{Alignment between Chinese companies' policies (DeepSeek and Baidu) and China's Generative AI Services measures. The figure compares the risk categories specified in the companies' policies with those outlined in the regulation at level-3 of our proposed taxonomy. The last row reports the overall agreement.}
    \label{fig:china_company}
\end{figure*}

Here we consider a case study of Chinese firms' policies and China's Interim Measures for the Management of Generative Artificial Intelligence Services.
As the US AI Executive Order largely imposes voluntary requirements and the EU AI Act is yet to take full effect, China's recent AI regulation, the Interim Measures for the Management of Generative Artificial Intelligence Services \cite{china-genai}, is perhaps the most impactful AI regulation currently in effect.
We use this regulation (specifically the 20 risk categories mapped to our taxonomy) and the policies of companies providing services within China (DeepSeek and Baidu) as a case study to analyze the alignment between the legally mandated risk categories and those specified in companies' policies. 
Figure \ref{fig:china_company} presents the results at level-3 of our taxonomy. The last row reports the overall degree of alignment in terms of the overlapping aspects of risks specified by company policies.

Our analysis shows that both companies' policies cover more than 90\% of the risk categories listed in the Generative AI Services measures. 
The only risk categories that are not referenced in both companies' policies are ``Autonomous Unsafe Operation of Systems'' and ``Advice in Heavily Regulated Industries,'' both from the ``Operational Misuse'' category. 
The law itself specifies ``Utilizing generative AI in high-security service areas (such as automated control systems, medical information services, psychological counseling, and critical information infrastructure)'' as a key risk with respect to generative AI services. 
Although the two companies do not explicitly mention these risk categories in their policies, they do allocate liability in their disclaimers \cite{Baidu_user, DeepSeek_user}, stating that users shall ``bear all risks associated with using this Service and its related content, including the truthfulness, completeness, accuracy, and timeliness of this Service and its content.''


\subsection{Takeaways}

We present three takeaways from this work:
\begin{enumerate}
    \item Including a larger number of categories in taxonomies of the risks posed by AI can be highly useful. By constructing a risk taxonomy with hundreds of categories, we provide a level of granularity that may be assist policymakers or industry policy researchers when drafting future AI policies. Without a greater level of detail in discussions of AI risk, it is difficult to understand that superficial alignment between policies on level-2 risk categories may not be reflective of any consistency in more specific level-4 risks. Many AI risk taxonomies include fewer than 50 risk categories and would benefit from greater depth.
    \item Government AI regulation may not be as expansive as is commonly claimed. As \citep{bommasani2024foundation} find, a close reading of the EU AI Act and the US AI Executive Order show that there are relatively few requirements for foundation model developers. We similarly find that the EU, US, and China include fewer risk categories in their regulations than AI companies have in their policies. As a result, governments may have room to enact additional requirements related to risk mitigation without imposing additional compliance burdens on some companies.
    \item Considering initiatives from a variety of different jurisdictions can significantly enhance analysis of AI safety \citep{Bradford2023, ConcordiaAI2023}. By including both regulations and policies from the US, EU, and China, we were better able to assess the regulatory environment facing multinational companies and potential opportunities for global cooperation on AI safety.\footnote{While we also consider policies from Cohere, which is based in Canada, we do not examine Canadian government regulations in this work, in part because the Artificial Intelligence and Data Act is still under development. In this work, we consider Cohere's policies in the context of its peers that also operate in the US.} We hope to analyze policies from a larger number of countries in future work. 
\end{enumerate}

\section{Conclusion}

In this work we construct a comprehensive risk taxonomy based on public and private sector policies that describe how governments and companies regulate risky uses of generative AI models. 
This method allows us to ground the AIR 2024 in existing practices, potentially making it a more tractable framework for risk mitigation.
We find substantial differences across companies and different kinds of company policies in terms of prohibited categories of risk, illustrating how different organizations conceptualize risks.
The union of risk categories contained in company policies is broader than that of existing government policies, showing that a lack of specificity in AI regulation may create gaps in enforcement.
We hope that this work can tangibly contribute to AI safety by serving as the basis for improved policies, regulations, and benchmarks. 

\newpage
\bibliographystyle{plain}
\bibliography{bibtex}

\end{document}